\documentclass[amsmath, amssymb, twocolumn, 14]{revtex4-1} 
\usepackage{appendix}
\usepackage[utf8]{inputenc}
\usepackage[section]{placeins}
\usepackage[dvipsnames]{xcolor}
\usepackage{bbold}
\usepackage{graphicx}
\usepackage{dsfont}
\usepackage[hidelinks]{hyperref}
\usepackage{amsmath}
\usepackage{wrapfig}

\usepackage{booktabs}
\AtBeginDocument{
  \heavyrulewidth=.08em
  \lightrulewidth=.05em
  \cmidrulewidth=.03em
  \belowrulesep=.65ex
  \belowbottomsep=0pt
  \aboverulesep=.4ex
  \abovetopsep=0pt
  \cmidrulesep=\doublerulesep   
  \cmidrulekern=.5em
  \defaultaddspace=.5em
}

\usepackage{braket}
\usepackage{multirow}
\usepackage{tabularx}

\usepackage{glossaries}
\newacronym{hf}{HF}{Hartree-Fock} 
\newacronym{vasp}{VASP}{Vienna ab initio simulation package}
\newacronym{ltmp2}{LTMP2}{Laplace transformed MP2}
\newacronym{diis}{DIIS}{direct inversion of the iterative subspace}
\newacronym{uhf}{UHF}{unrestricted \gls{hf}} 
\newacronym{mp2}{MP2}{second-order M\o ller-Plesset perturbation theory}
\newacronym{ump2}{UMP2}{unrestricted \gls{mp2}}
\newacronym{CIS}{CIS}{configuration interaction singles}
\newacronym{RPA}{RPA}{random phase approximation}
\newacronym{ccsd}{CCSD}{coupled cluster singles doubles}
\newacronym{ccsdt}{CCSD(T)}{coupled cluster singles doubles and perturbative triples} 
\newacronym{uccsd}{UCCSD}{unrestricted \gls{ccsd}}
\newacronym{dft}{DFT}{density functional theory}
\newacronym{tddft}{TD-DFT}{time-dependent density functional theory}
\newacronym{cc}{CC}{coupled cluster}
\newacronym{fci}{FCI}{Full Configuration Interaction}
\newacronym{eom}{EOM}{equation of motion}
\newacronym{eomcc}{EOM-CC}{equation of motion coupled cluster}
\newacronym{ipeomcc}{IP-EOM-CC}{Ionization Potential \gls{eomcc}}
\newacronym{eaeomcc}{EA-EOM-CC}{Electron Attachment \gls{eomcc}}
\newacronym{eeeomcc}{EE-EOM-CC}{Electron Excitation \gls{eomcc}}
\newacronym{eeeomccsd}{EE-EOM-CCSD}{Equation of motion \gls{ccsd}}
\newacronym{cc4s}{\texttt{Cc4s}}{coupled cluster for solids}
\newacronym{ctf}{\texttt{CTF}}{Cyclops Tensor Framework}
\newacronym{bse}{BSE}{Bethe-Salpeter equation}
\newacronym{qd}{QD}{quantum dot}
\newacronym{CBS}{CBS}{complete basis set}
\newacronym{Qmc}{QMC}{Quantum Monte Carlo}
\newacronym{qmc}{QMC}{quantum Monte Carlo}
\newacronym{fno}{FNO}{frozen natural orbital}
\newacronym{fnos}{FNOs}{frozen natural orbitals}
\newacronym{bsie}{BSIE}{basis set incompleteness error}
\newacronym{fsie}{FSIE}{finite size error}

\def\SupportingInformationHfCcsdEnergiesTables{I}
\def\SupportingInformationTwistAveragingTables{II-VII} 

\begin{document}

\title{Formation energies of silicon self-interstitials using periodic coupled cluster theory}

\author{Faruk Salihbegovi\'c}
\author{Alejandro Gallo}
\author{Andreas Grüneis}
\affiliation{Institute for Theoretical Physics,
 Vienna University of Technology (TU Wien).
 A-1040 Vienna, Austria, EU.}

\date{\today}

\begin{abstract}
We present a study of the self-interstitial point defect formation
energies in silicon using a range of quantum chemical theories
including the \gls{cc} method within a periodic supercell approach.
We study the formation energies of the X, T, H and C3V
self-interstitials and the vacancy V.
Our results are compared to findings obtained using different ab
initio methods published in the literature and partly to experimental
data.
In order to achieve computational results that are converged with
respect to system size and basis set, we employ the recently proposed
finite size error corrections and basis set incompleteness error
corrections.
Our CCSD(T) calculations yield an order of stability of the X, H and T
self-interstitials, which agrees both with quantum Monte Carlo results
and with predictions obtained using the random-phase approximation
as well as using screened hybrid functionals.
Compared to quantum Monte Carlo results with backflow corrections, the
CCSD(T) formation energies of X and H are only slightly larger by
about $100\,$meV.
However, in the case of the T self-interstitial, we find significant
disagreement with all other theoretical predictions. Compared to
quantum Monte Carlo calculations, CCSD(T) overestimates the formation
energy of the T  self-interstitial by $1.2\,$eV.
Although this can partly be attributed to strong correlation effects,
more accurate electronic structure theories are needed to understand
these findings.
\end{abstract}

\maketitle

\section{Introduction}
After more than 60 years of sophisticated silicon device production, one might think that all the details of this material are fully understood, especially knowing that the manufacturing of today's nanometer-sized transistors requires near-atomic accuracy.
However, as a direct result of this miniaturization, the accidental creation of a single trapping center can be large enough to alter the electronic properties of the sample, making this issue the most feared phenomenon in the industry~\cite{DeviceFabrication}.

To better understand the influence of single isolated vacancies and interstitials, these have to be produced experimentally.
This can be achieved with 1--3$\,$MeV electron irradiation performed at cryogenic temperatures.
The identification of these centers is possible through characterization techniques like electron paramagnetic spectroscopy (EPR),
which is capable of targeting the atomic distortion triggered by the form of the localized electronic density~\cite{EPR,EPR2}.
Infrared optical absorption and deep-level transient spectroscopy can also be used to identify
center-induced states within the semiconductor gap~\cite{Last40Years,DLTS,OAS}.
The availability of experimental data motivated the development of
simple theoretical models geared towards quantitatively reproducing
the basic features of these defects.
Furthermore, the rapid growth in computational resources made it possible to perform ab initio calculations to
model and understand their properties thoroughly on an atomistic level.

Point defects, such as vacancies, interstitials and anti-site defects, are the only thermodynamically stable defects at finite temperatures~\cite{ThermodynamicStability}.
The presence of point defects often controls the kinetics of the material and can therefore fundamentally alter its electronic, optical and mechanical properties.
This makes the understanding of point defects technologically important for a wide range of applications such as doping of semiconductors~\cite{Doping1,Doping2,Doping3,Doping4}, production of quantum devices~\cite{qd1,qd2}, and controlling the transition temperature of shape memory alloys~\cite{ShapeMemory}.

Of all materials, silicon is one of the most important for industrial use and plays a crucial role in a wide variety of devices, e.g., advanced electronic devices, power devices, solar cells, and microelectronic systems.
In all these applications, Czochralski (CZ) and floating zone (FZ) silicon single crystals are used (except for some solar cells)~\cite{experimentalData}.
The diffusion characteristics and thermodynamics of silicon self-interstitials and vacancies dominate the doping and annealing processes for electronics applications~\cite{Doping2,Doping3}.
However, the understanding of self-diffusion in silicon remains incomplete despite decades of research~\cite{dow5,dow6,dow7,dow8,dow9,dow10,dow11,dow12,dow13,dow14,dow15,dow16,dow17,dow18,dow19,dow20,dow21,dow22,dow23,exp6,qd1,QMCRef}.
Questions regarding the role of the self-interstitials and the vacancy in the self-diffusion remain.
One of the remaining questions, which needs to be addressed using quantum mechanical methods,
is the formation energy of the silicon self-interstitials and vacancy.
The most widely used method in this regard is \gls{dft}, which replaces the complicated many-body electron interactions with quasi particles interacting via an exchange and correlation functional.
Exchange and correlation functionals based on the local density approximation (LDA), general gradient approximation (GGA) and hybrid functionals
predict formation energies in a range of $2\,$eV-$4.5\,$eV~\cite{dow10}.
Green's function based methods, such as the $GW$ approximation, are expected to yield more accurate results and
predict formation energies of about $4.5\,$eV~\cite{dow7}.
A low-scaling implementation of the random-phase approximation reported formation energies
on a similar scale~\cite{Kresse}.
\gls{Qmc} provides another computationally more expensive alternative to \gls{dft} and is among
the most accurate electronic structure methods available.
Several groups have calculated the formation energies using \gls{qmc}~\cite{QMCOld,QMCRef}.
In this work, we focus only on the former~\cite{QMCRef}, because it employs a Slater-Jastrow-backflow wavefunction, which changes the formation energies substantially.

The \gls{cc} method is a systematically improvable many-electron theory, which
is widely used in molecular quantum chemistry, where it achieves high accuracy in the prediction of reaction energies for a wide range of systems.
While being an efficient method for calculating small to medium-sized molecules,
single-reference \gls{cc} methods have never been used to calculate the formation energies of
silicon self-interstitials and the vacancy in diamond cubic crystal silicon.
Only over the past few years have computationally efficient implementations
of periodic \gls{cc} methods become available to study such systems~\cite{Booth2013,gruber18b,Gruber2018,mcclain_2017}.
Moreover, recent developments in embedding approaches also make it possible to study such local phenomena using
\gls{cc} methods~\cite{Schutz2017,Usvyat2018,Nusspickel2021,Chen2020,Sauer2019,Lin2020,Goodpaster2014,Schaefer2021a,Lau2021}.
The goal of this work is to calculate the formation energies of the silicon self-interstitials and vacancy at the level of \gls{ccsdt} theory and compare them to experimental data~\cite{exp2,exp4,exp5,exp6,exp7} and reference data from literature~\cite{QMCRef,Kresse, dow7, dow10}.

\section{Theory and Methods}\label{theory}
\subsection{\gls{hf} and \gls{cc} theory}\label{HFBasicTheory}
In \gls{hf} theory, the many-body wavefunction is approximated by a single Slater determinant, and the energy is optimized with respect to variations of the spin orbitals used to construct the Slater determinant. The Slater determinant constructed from these spin orbitals is the \gls{hf} ground state wavefunction $\ket{0}$ and can be interpreted as a new vacuum from which particle-hole pair excitations
are created and annihilated in the context of quantum field theory.
Building on one-body theories such as \gls{hf}, \gls{cc} theory employs an exponential ansatz acting on a single Slater
determinant.
\begin{gather}
  \ket{\Psi_{\mathrm{CC}}}
    = e^{\hat{T}}\ket{0}%
    ,\\
  \hat{T}
    = \sum_{i,a}
        t_i^a \hat{a}_{a}^{\dagger} \hat{a}_i
      + \frac{1}{4}\sum_{i,j,a,b} t_{ij}^{ab}
          \hat{a}_{a}^{\dagger} \hat{a}_{b}^{\dagger} \hat{a}_j\hat{a}_i
      + \cdots
\end{gather}
Herein, the indices $i$, $j$ and $a$, $b$ refer to particle (unoccupied) and hole (occupied) states, respectively.
By projecting onto the excited Slater determinants, one obtains a set of coupled non-linear equations,
that can be solved for the amplitudes $t_{i}^{a}$, $t_{ij}^{ab}$, etc. using iterative methods.
For practical calculations, the cluster operator $\hat{T}$ has to be truncated, usually to single and double excitations.
While including the full triple excitation manifold is computationally too expensive, an estimate of the connected triples contribution can be calculated noniteratively using an expression reminiscent of many-body perturbation theory.
This method is referred to as \gls{ccsdt} theory~\cite{A.fifth.order.pRaghav1989}.
A more detailed description of \gls{cc} methods can be found in Ref.~\cite{bartlett,book:293288}.

\subsection{Basis set and finite size error correction}\label{bsec}
All practical post-Hartree-Fock calculations of real materials
employ a finite number of particle states also referred to as virtual orbitals $N_\mathrm{v}$.
The truncation of the virtual orbital basis set introduces the \gls{bsie}.
The \gls{bsie} vanishes very slowly in the limit of $N_\mathrm{v}\rightarrow\infty$.
This observation is not surprising since the leading order basis set error
originates from the so-called electron-electron cusp conditions~\cite{On.the.eigenfunKato.1957},
which are in real space a short-ranged electronic correlation phenomenon.
Explicitly correlated methods help reduce the finite basis
set error substantially~\cite{G17,G18}.
These methods, as well as their application to periodic systems, have already been discussed extensively elsewhere~\cite{G11,G19,G20}.
In order to treat the \gls{bsie} we use a pair-specific cusp correction for \gls{cc} theory~\cite{FocalPoint}.
This scheme is based on \gls{fnos} and diagrammatically decomposed contributions to the electronic correlation energy, which dominates the \gls{bsie}.
To partly account for the \gls{bsie} of the (T) contribution to the CCSD(T) correlation energy,
we rescale the (T) contribution using the ratio of the CCSD correlation energy with and without the \gls{bsie} correction
discussed above.

Further, we simulate the silicon crystal using a periodic supercell approach with a finite system size.
The employed finite system introduces the \gls{fsie}.
It should be noted that many properties, including the ground state energy, converge slowly with respect to the system size.
This originates from the fact that correlated wavefunction-based theories capture longer ranged electronic correlation effects such as dispersion interaction explicitly.
The coupled cluster correlation energy can be expressed as an integral over the electronic transition structure factor
multiplied by the Coulomb kernel in reciprocal space.
Finite size errors partly originate from an incomplete sampling of this integral in reciprocal space.
Here, we employ an interpolation technique that can be used to evaluate the correlation energy
integral more accurately, reducing the finite size error.
The technical details are described in reference~\cite{GrueneisFiniteSize}.

\subsection{Cell structure}\label{CellStructure}
The employed simulation cells of silicon self-interstitials are obtained by adding one Si atom to
the diamond cubic crystal structure of bulk silicon and relaxing the atomic positions.
The energetically most stable silicon self-interstitial (X) is one in which two silicon atoms
reside symmetrically shifted from the position previously occupied by one.
The two atoms are oriented parallel to the [110] direction.
The second most favorable self-interstitial (H) is  where the additional Si atom
is equidistant to six other atoms, forming a hexagonal ring.
It is worth noting that this configuration is unstable in DFT-PBE, where the central atom of
the ring is slightly moving away in a direction orthogonal to the ring (C3V)~\cite{C3V1,dow7}.
The last self-interstitial considered in this work, with the highest energy, is where the additional Si
atom is coordinated equidistantly to four nearest neighbors, forming a tetrahedron (T).
For the T interstitial, the highest occupied state is threefold degenerate but only occupied by two electrons, which potentially introduces a multireference character.
The vacancy is created by removing one Si atom from the bulk structure.
Like the T interstitial, it has a threefold degenerate highest occupied state, occupied by two electrons.
The vacancy is known to undergo a Jahn-Teller distortion to a $D_{2d}$ symmetry~\cite{JahnTeller}.

All the ion positions of the used structures have been relaxed at the DFT level
using the  PBE exchange and correlation energy functional~\cite{PBE}.
The shape and volume of the cells are kept fixed during the relaxation procedure.

\subsection{Computational Details}\label{ComputationalDetails}

All coupled-cluster calculations have been performed using our high-performance open-source
coupled cluster simulation code, \gls{cc4s}.
The preparation of the necessary reference wavefunction and the required intermediates was performed
using the \gls{vasp}~\cite{vasp1,vasp2,vasp3}.
For all calculations in \gls{vasp} a plane-wave kinetic energy cut off of $E_{\text{cut}}=400\,$eV has been used.
The employed smearing parameter is $\sigma=10^{-4}\,$eV and a convergence criterion of $\Delta E=10^{-6}\,$eV
is used.
All other numerical parameters were left unchanged from their default values.
The \gls{hf} calculations are performed using  \gls{vasp} and the
structures described in section~\ref{CellStructure}.
The \gls{hf} calculations are done using a $\Gamma$-centered $7\times7\times7$ $k$-point mesh.
All post-HF calculations sample the first Brillouin zone using a single  $k$-point only.
Further, we need to compute all unoccupied \gls{hf} orbitals since in \gls{cc} theory we approximate the many-electron wavefunction using excited Slater determinants, constructed by occupied and unoccupied \gls{hf} orbitals.
In \gls{vasp} this is achieved by setting the number of virtual orbitals to the maximum number of plane-waves in the basis set.
The convergence of the \gls{ccsd} electron correlation energy is very slow when using canonical \gls{hf} orbitals.
A much faster convergence to the complete basis set limit is achieved using natural orbitals.
In \gls{vasp} approximate natural orbitals can be calculated as described in Eq.(2) from Ref.~\cite{MP2NO}.
After calculating all natural orbitals, a subset of them is chosen for the \gls{cc4s} calculations.
For the coupled-cluster theory calculations, we chose the number of unoccupied natural orbitals per
occupied orbital to be 5, 10, 15, 20, 25 and 30.
Additionally, for the basis set correction algorithm described in section~\ref{bsec} and in the references therein, the \gls{mp2} pair energies are needed.
For this purpose, there are two algorithms available in \gls{vasp}~\cite{LTMP2,Marsman2009}.
In our case, we have used a 16-atom cell for the bulk with 32 occupied orbitals; therefore, the \gls{mp2} algorithm from Ref.~\cite{Marsman2009}
is more efficient.
In the case of more than 50 occupied orbitals, a different algorithm based on \gls{ltmp2} might be faster and less memory consuming~\cite{LTMP2}.
Note that the basis set correction algorithm uses a focal-point approach, and from now on, the basis set correction
is also referred to as the focal-point correction (FPC).
With these preparations done, \gls{vasp} can provide all necessary files needed for the \gls{ccsdt} calculation
with the finite size and basis set error correction computed by \gls{cc4s}.
It is worth noting that the \gls{ccsd} calculation in \gls{cc4s} converges much faster when using
the \gls{diis} mixer instead of the default linear mixer.
The described workflow with all necessary files can be found on GitHub or zenodo~\cite{github,zenodo}.

After studying the \gls{bsie}, we chose our number of virtual orbitals per occupied orbital to be 10 and repeated all calculations with 10 randomly chosen $k$-point shifts in order to get a twist average estimate of the \gls{ccsdt} correlation energies.

All calculations have been performed using 16 compute nodes, each equipped with 384~GB main memory.

\section{Results}
We now discuss the formation energies of the silicon self-interstitial structures described in
section~\ref{CellStructure} at the \gls{ccsdt} level of theory.
We use 16 atom cells for the pristine bulk crystal with periodic boundary conditions, the
interstitial cells have 17 atoms, while the vacancy has 15 atoms.
The \gls{hf} energies, \gls{ccsd}, \gls{ccsdt}, finite size and basis set energy corrections
can be found in the supplementary material, in table~\SupportingInformationHfCcsdEnergiesTables{}.
The formation energy is calculated by subtracting the energy of the
interstitial cell with the energy of the bulk cell scaled to the same
number of atoms:
\begin{align}
E_{\text{F}} = E_{\text{int}}-\frac{N_{\text{int}}}{N_{\text{bulk}}}E_{\text{bulk}}\label{formationequ}.
\end{align}
\begin{figure}[htb]
  \centering
  \includegraphics[]{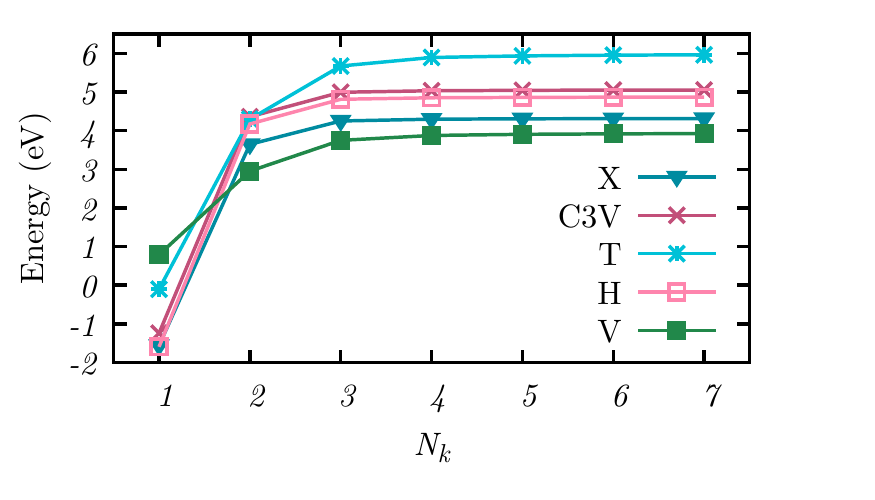}
  \caption{%
    The formation energy as a function of the number of $k$-points
    used in the Hartree-Fock calculation for all self-interstitials.
    A $\Gamma$-centered cubic $k$-point mesh was used with $N_k$ x
    $N_k$ x $N_k$ gridpoints.
    \label{fig:kpointConvergence}}
\end{figure}
We first discuss the convergence of the  \gls{hf} energy contribution to the formation energies.
Figure~\ref{fig:kpointConvergence} shows the convergence of the \gls{hf} formation energies
with respect to the size of the $k$-point mesh used in the \gls{hf} calculation.
We used a $\Gamma$-centered cubic $k$-point mesh with
up to 7$\times$7$\times$7 grid points.
Using only one $k$-point gives qualitatively and quantitatively wrong formation energies.
With a $k$-point mesh size of 5$\times$5$\times$5, the formation energies are already well converged and increasing the $k$-point grid size further to 7$\times$7$\times$7 increases the formation energies of X, H and C3V by less than $8.5\,$meV, while T and V increase by $29\,$meV and $19\,$meV.
\begin{figure}[htb]
  \centering
  \includegraphics[]{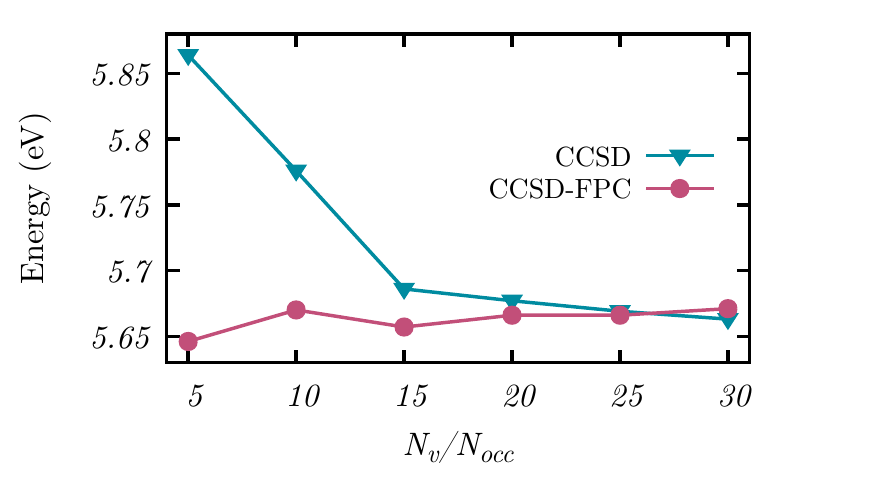}
  \caption{%
    \gls{ccsd} formation energy of the X interstitial as a function
    of the number of orbitals per occupied orbital with and without the
    basis set correction scheme (FPC).
    A $\Gamma$-centered cubic mesh was used.}%
    \label{fig:basisSetConvergence}
\end{figure}
We now discuss the convergence of the correlation energy contributions to the formation energies
with respect to the number of virtual orbitals. Let us note that the \gls{hf} formation energy
Contributions are independent of the virtual orbital basis set size.
The MP2 calculations employ the complete virtual orbital basis set defined by the kinetic energy cutoff of the plane wave basis set. Furthermore the MP2 correlation energies are automatically extrapolated to the complete basis set
limit using a procedure explained in Ref.~\cite{LTMP2}.
For the post-MP2 correlation energy calculations, we employ approximate natural orbitals as virtual orbitals
and  seek to converge the correlation energies explicitly by increasing
the number of virtual orbitals. We find that this approach allows for an effective cancellation between
basis set incompleteness errors of correlation energies for different systems when taking their differences.
Furthermore, we add a basis set incompleteness error correction described in Ref.~\cite{FocalPoint}
to accelerate the convergence of the CCSD correlation energy.
The effect of the basis set correction on the formation energy is highlighted in figure~\ref{fig:basisSetConvergence},
which depicts the formation energy retrieved as a function of the number of virtual orbitals per occupied orbital for the X interstitial.
Our results indicate that between 10 and 20 virtual orbitals per occupied orbital suffice to achieve converged formation energies with and without the basis set correction, respectively. The remaining basis set incompleteness error is caused by fluctuations
on the scale of about $10\,$meV, which is smaller than the expected accuracy of the employed theories.
\begin{figure}[htb]
\begin{center}
\includegraphics[]{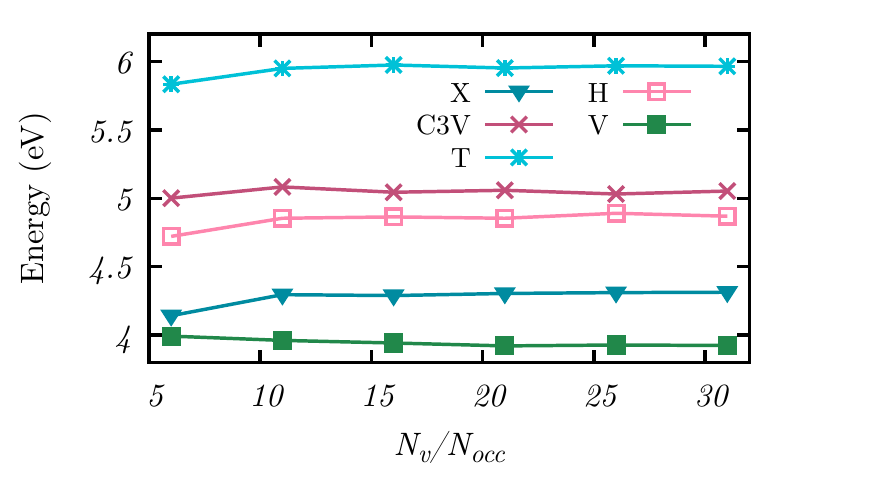}
\caption{%
  \gls{ccsdt} formation energies as a function of the number of virtual orbitals
  per occupied orbital for all
  self-interstitials include finite size and basis set corrections.}%
  \label{fig:orbitalConvergence}
\end{center}
\end{figure}
Figure~\ref{fig:orbitalConvergence} shows the convergence of the \gls{ccsdt} formation energies of all self-interstitials with respect
to the number of natural orbitals, including the finite size and basis set corrections.
Note that the basis set correction behaves similarly for all self-interstitials.
From figure~\ref{fig:orbitalConvergence}, we see that
$N_\mathrm{v}/N_\mathrm{occ}=10$ is already accurate enough to assume
convergence within chemical accuracy ($\approx43\,$meV).

With $N_\mathrm{v}/N_\mathrm{occ}=10$, we repeat all calculations at 10 random $k$-points in order to obtain a twist-averaged
estimate of the correlation energy contribution to the formation energies.
This approach reduces finite size errors in CC calculations that originate from single-particle effects~\cite{Gruber2018}.
The energy corrections for the twist averaging can be found in the supplementary material, in tables~\SupportingInformationTwistAveragingTables{}.
Using 10 random $k$-points our standard deviation from the average \gls{ccsdt} energy is in decreasing order for $E_{\text{F}}(\text{T})=227\,$meV,
$E_{\text{F}}(\text{V})=101\,$meV, $E_{\text{F}}(\text{X})=80\,$meV, $E_{\text{F}}(\text{C3V})=73\,$meV and $E_{\text{F}}(\text{H})=51\,$meV.
After twist averaging, the formation energy of V, T and X increases by $818\,$meV, $367\,$meV and $239\,$meV.
While the formation energy of the C3V and H interstitial decreases by $85\,$meV and $44\,$meV,
indicating that it is important to account for this contribution.
\begin{table*}
\centering
\caption{\gls{ccsd} and \gls{ccsdt} formation energies of the silicon self-interstitials and the vacancy with and without the basis set and finite size correction as a function of the unoccupied to occupied orbital ratio $N_{\mathrm{v}}/N_{\mathrm{occ}}$ at the $\Gamma$-point.
FPC and FS denote that the basis set/finite size corrections are included.
\label{table2}}
\begin{ruledtabular}
\begin{tabular}{lccccccc}
\toprule
  \gls{hf}/\gls{mp2} & $N_{\mathrm{v}}/N_{\mathrm{occ}}$ & \gls{ccsd} & \gls{ccsdt} & \gls{ccsd}-FS & \gls{ccsd}-FPC & \gls{ccsd}-FS-FPC & \gls{ccsdt}-FS-FPC \\
\midrule
C3V     &5      &6.659  &5.981  &6.149          &6.334          &5.824         &5.001\\
8.502   &10     &6.484  &5.753  &6.012          &6.343          &5.871         &5.083\\
4.408   &15     &6.374  &5.614  &5.907          &6.305          &5.837         &5.044\\
        &20     &6.347  &5.577  &5.880          &6.317          &5.850         &5.059\\
        &25     &6.330  &5.556  &5.864          &6.288          &5.822         &5.031\\
        &30     &6.315  &5.537  &5.849          &6.308          &5.843         &5.053\\
\midrule
X       &5      &5.864  &5.094  &5.285          &5.646          &5.067         &4.142\\
7.930   &10     &5.776  &4.982  &5.254          &5.670          &5.148         &4.296\\
3.780   &15     &5.686  &4.866  &5.169          &5.657          &5.141         &4.289\\
        &20     &5.677  &4.853  &5.162          &5.666          &5.150         &4.305\\
        &25     &5.669  &4.842  &5.155          &5.666          &5.152         &4.311\\
        &30     &5.663  &4.831  &5.150          &5.671          &5.158         &4.313\\
\midrule
T       &5      &7.857  &7.051  &7.238          &7.435          &6.816         &5.833\\
9.954   &10     &7.628  &6.764  &7.055          &7.455          &6.882         &5.949\\
5.355   &15     &7.531  &6.641  &6.964          &7.469          &6.902         &5.974\\
        &20     &7.511  &6.611  &6.944          &7.447          &6.880         &5.952\\
        &25     &7.489  &6.584  &6.923          &7.458          &6.892         &5.968\\
        &30     &7.476  &6.565  &6.910          &7.456          &6.890         &5.964\\
\midrule
H       &5      &6.358  &5.717  &5.877          &5.986          &5.504         &4.722\\
8.162   &10     &6.231  &5.538  &5.783          &6.053          &5.605         &4.854\\
4.212   &15     &6.132  &5.413  &5.689          &6.058          &5.615         &4.864\\
        &20     &6.103  &5.376  &5.661          &6.046          &5.604         &4.854\\
        &25     &6.090  &5.360  &5.649          &6.075          &5.635         &4.890\\
        &30     &6.075  &5.341  &5.634          &6.056          &5.616         &4.869\\
\midrule
V       &5      &5.291  &4.866  &4.820          &4.999          &4.528         &3.993\\
5.554   &10     &5.034  &4.546  &4.573          &4.952          &4.492         &3.961\\
4.305   &15     &4.960  &4.443  &4.500          &4.942          &4.483         &3.942\\
        &20     &4.942  &4.413  &4.484          &4.925          &4.467         &3.921\\
        &25     &4.926  &4.394  &4.470          &4.928          &4.472         &3.928\\
        &30     &4.919  &4.380  &4.462          &4.928          &4.472         &3.924\\
\bottomrule
\end{tabular}
\end{ruledtabular}
\end{table*}

For completeness, we summarize the obtained formation energies at the different levels of theory again in table~\ref{table2}.
The \gls{hf} energies, \gls{ccsd}, \gls{ccsdt}, finite size and basis set energy corrections
can be found in the supplementary material, in table~\SupportingInformationHfCcsdEnergiesTables{}.

Next, we briefly discuss the effect of the \gls{fsie} correction based on the structure factor interpolation,
which accounts for two-electron finite size errors.
Table~\ref{table2} summarizes the computed \gls{ccsd} formation energies with and without
the corresponding finite size correction denoted as \gls{ccsd}-FS and \gls{ccsd}, respectively.
It is not surprising that this correction is significant and on the scale of about $0.5\,$eV.
However, based on previous results reported in Ref.~\cite{Gruber2018}
we expect that the employed finite size correction will suffice for the 16/17 atom cells to achieve chemical accuracy in the convergence of the computed formation energies with respect to the employed system size.
Furthermore, it can be concluded from the comparison between \gls{ccsd}-FS and \gls{ccsd} in table~\ref{table2},
that the computed finite size correction is already well converged using $N_{\mathrm{v}}/N_{\mathrm{occ}}=10$.
Note that the finite size correction can currently only be applied to the CCSD calculation. The (T) contribution to the formation energies
is significantly smaller than the CCSD correlation energy contribution, which makes it plausible to neglect the finite size correction
to the (T) contribution.

Table~\ref{table2} also includes results for the vacancy formation energy. We note that these calculations employ
a 15-atom cell only. Due to the small supercell size, the system does not undergo a  Jahn-Teller distortion~\cite{JahnTeller},
which can be observed for larger cells and which significantly changes the formation energy. Therefore, we note that these results are only meaningful as benchmarks for other theories employing identical geometries and can not be compared to experiment.
\begin{table*}
\centering
        \caption{Computed and converged \gls{hf}, \gls{ccsd} and \gls{ccsdt} formation energies including
        all reported corrections in this work compared to QMC~\cite{QMCRef}, RPA~\cite{Kresse}, PBE~\cite{Kresse},
        LDA~\cite{dow7,dow10}, $G_0W_0$~\cite{dow7} and HSE~\cite{dow10} from the literature
        and also experimental data~\cite{exp2,exp4,exp5,exp6,exp7}. All results have been obtained for the 16/17 atom cells
        except RPA(216), which employed 216/217 atom cells.
        \label{table4}}
\begin{ruledtabular}
\begin{tabular}{lcccccccccccc}
\toprule
Cell            &\gls{hf}       &\gls{ccsd}     &\gls{ccsdt}    &QMC    &QMC (nobf)     &$G_0W_0$&RPA   &RPA (216)      &HSE    &PBE    &LDA    &Exp.           \\
\midrule
X               &7.930          &5.295          &4.535          &4.4    &4.9            &4.46   &4.27   &4.2            &4.46   &3.56   &3.29   &               \\
T               &9.954          &7.127          &6.316          &5.1    &5.2            &       &4.53   &4.93           &4.92   &3.66   &3.56   &               \\
H               &8.162          &5.559          &4.810          &4.7    &4.9            &4.4    &4.45   &4.33           &4.82   &3.74   &3.4    &4.2 – 4.7      \\
\bottomrule
\end{tabular}
\end{ruledtabular}
\end{table*}
Our best estimates of the formation energies at the level of \gls{hf}, \gls{ccsd} and  \gls{ccsdt} theory
including all corrections discussed above
are compared to values from the literature and experiment in table~\ref{table4}.
We see that the formation energies calculated with LDA and PBE are small (3--4$\,$eV) and close to each other for all self-interstitials, with a difference of $80-160\,$meV.
Yet the order of stability is not the same; for LDA the most stable self-interstitial is X then H and T.
For PBE T has a lower energy than H.
Incorporating a portion of the exact exchange correlation energy in the HSE functional increases the formation energies,
and their difference to $360\,$meV and $100\,$meV, with the order of stability of X, H and T.
RPA predicts the same order of stability with a difference between X and H of $180\,$meV, while the difference between H and T is $80\,$meV.
Increasing the cell size to 216 atoms changes the differences significantly to $130\,$meV and $600\,$meV, still with the same order of stability.
$G_{0}W_{0}$ for 16 atom cells predicts that H is more stable than X while their difference is only $60\,$meV.
In \gls{qmc}, using 16 atom cells without a backflow correction, X and H are nearly degenerate, while the difference to T is $300\,$meV.
Including the backflow correction gives the order of stability as X, H and T with clear differences of $300\,$meV and $400\,$meV.
We now turn to the wavefunction methods employed in this work.
The formation energies calculated with \gls{hf} are much larger than the ones calculated with the other theories presented.
While the order of stability is in agreement with the corrected \gls{qmc} calculations, their difference is $232\,$meV and $1792\,$meV.
Expanding the correlation space further to \gls{ccsd} and \gls{ccsdt} theory, including the basis set and finite size correction, lowers the
formation energies by $2.8$--$2.6\,$eV and another $811$--$749\,$meV.
Their relative difference also changes to $264\,$meV and $1568\,$meV for \gls{ccsd} theory and $275\,$meV and $1506\,$meV for \gls{ccsdt} theory.
Our estimated \gls{ccsdt} formation energies are in good agreement with extrapolated \gls{qmc} calculations~\cite{QMCRef} employing a
Slater-Jastrow-backflow correction for the X and H interstitial.
However, we have a discrepancy of $1.2\,$eV for the T interstitial.
This could stem from the fact that in \gls{dft} the energetically highest occupied orbitals are threefold degenerate while being occupied by two electrons. It may be an indication that a multireference treatment is needed.
Our \gls{ccsdt} formation energy for the H interstitial is within reasonable agreement with experiment, being $110\,$meV
above the experimental upper bound.

\section{Conclusion and Summary}
In this paper, we have calculated the formation energies of the silicon self-interstitials and the vacancy in a periodic supercell at the
\gls{ccsdt} level of theory. We have used correction schemes tailored to \gls{cc} theory to reduce the \gls{bsie} and the \gls{fsie}.
Our results have been compared to data from the literature and experiment, including LDA, PBE, HSE, RPA, $G_0W_0$ and \gls{qmc}.

In general, \gls{dft} using the LDA and PBE functionals fails to differentiate the structures, resulting in
small energy differences between the self-interstitials while also underestimating the formation energies.
Additionally, \gls{hf} overestimates the formation energies.
The HSE functional offers a compromise, and its formation energies are in good agreement with the much more expensive
and accurate \gls{qmc} calculations.
The \gls{qmc} formation energies of the two most stable self-interstitials, X and H, are nearly degenerate.
This degeneracy is lifted by employing the Slater-Jastrow-backflow trial wavefunction~\cite{QMCRef}.

Our \gls{ccsdt} formation energies are in good agreement with
\gls{qmc} calculations employing a Slater-Jastrow-backflow
wavefunction for the X and H interstitials.
The \gls{ccsdt} formation energy for the H interstitial is within reasonable agreement with experimental data, being $110\,$meV
above the upper bound.
However, the \gls{ccsdt} formation energy of the T interstitial is $1.2\,$eV higher than in the \gls{qmc} calculations.
Since in \gls{dft} the highest occupied orbital of the T interstitial is threefold degenerate but only occupied by two electrons, we suppose a multireference approach may be necessary.
We stress that none of the discussed methods is expected to work for strongly correlated systems. DFT based approaches underestimate
the formation energy of strongly correlated defects due to the introduction of partly filled orbitals that reduce the self-interaction error.
QMC techniques require multideterminant trial wavefunctions for strongly correlated
systems to reduce the error from the fixed-node approximations, and RPA is expected to inherit part of the DFT errors for the treatment of strongly correlated systems.
Therefore, we have to conclude that more sophisticated theories will be needed in future studies to fully resolve the observed discrepancy for
the formation energy of the T interstitial.

Although we demonstrated that basis set convergence can be achieved efficiently at the level of \gls{ccsdt} theory using recently presented methods, the treatment of finite size errors is still challenging and relatively large defect concentrations had to be employed.
However, we note that recently developed embedding methods will allow to investigate much lower defect concentrations in a computationally efficient manner~\cite{embedding}.

\section*{Acknowledgements}
The authors thankfully acknowledge support and funding from the European Research Council (ERC) under the European Union's Horizon 2020 research and innovation program (Grant Agreement No 715594).
The computational results presented have been achieved using the Vienna Scientific Cluster (VSC).

\bibliography{Paper.bib}

\begin{thebibliography}{78}%
\makeatletter
\providecommand \@ifxundefined [1]{%
 \@ifx{#1\undefined}
}%
\providecommand \@ifnum [1]{%
 \ifnum #1\expandafter \@firstoftwo
 \else \expandafter \@secondoftwo
 \fi
}%
\providecommand \@ifx [1]{%
 \ifx #1\expandafter \@firstoftwo
 \else \expandafter \@secondoftwo
 \fi
}%
\providecommand \natexlab [1]{#1}%
\providecommand \enquote  [1]{``#1''}%
\providecommand \bibnamefont  [1]{#1}%
\providecommand \bibfnamefont [1]{#1}%
\providecommand \citenamefont [1]{#1}%
\providecommand \href@noop [0]{\@secondoftwo}%
\providecommand \href [0]{\begingroup \@sanitize@url \@href}%
\providecommand \@href[1]{\@@startlink{#1}\@@href}%
\providecommand \@@href[1]{\endgroup#1\@@endlink}%
\providecommand \@sanitize@url [0]{\catcode `\\12\catcode `\$12\catcode
  `\&12\catcode `\#12\catcode `\^12\catcode `\_12\catcode `\%12\relax}%
\providecommand \@@startlink[1]{}%
\providecommand \@@endlink[0]{}%
\providecommand \url  [0]{\begingroup\@sanitize@url \@url }%
\providecommand \@url [1]{\endgroup\@href {#1}{\urlprefix }}%
\providecommand \urlprefix  [0]{URL }%
\providecommand \Eprint [0]{\href }%
\providecommand \doibase [0]{http://dx.doi.org/}%
\providecommand \selectlanguage [0]{\@gobble}%
\providecommand \bibinfo  [0]{\@secondoftwo}%
\providecommand \bibfield  [0]{\@secondoftwo}%
\providecommand \translation [1]{[#1]}%
\providecommand \BibitemOpen [0]{}%
\providecommand \bibitemStop [0]{}%
\providecommand \bibitemNoStop [0]{.\EOS\space}%
\providecommand \EOS [0]{\spacefactor3000\relax}%
\providecommand \BibitemShut  [1]{\csname bibitem#1\endcsname}%
\let\auto@bib@innerbib\@empty
\bibitem [{\citenamefont {Graff}(2013)}]{DeviceFabrication}%
  \BibitemOpen
  \bibfield  {author} {\bibinfo {author} {\bibfnamefont {K.}~\bibnamefont
  {Graff}},\ }\href
  {https://www.researchgate.net/publication/316792450_Metal_Impurities_in_Silicon-Device_Fabrication}
  {\emph {\bibinfo {title} {Metal impurities in silicon-device fabrication}}},\
  Vol.~\bibinfo {volume} {24}\ (\bibinfo  {publisher} {Springer Science \&
  Business Media},\ \bibinfo {year} {2013})\BibitemShut {NoStop}%
\bibitem [{\citenamefont {Weber}(1981)}]{EPR}%
  \BibitemOpen
  \bibfield  {author} {\bibinfo {author} {\bibfnamefont {E.}~\bibnamefont
  {Weber}},\ }\href {\doibase https://doi.org/10.1002/crat.19810160215}
  {\bibfield  {journal} {\bibinfo  {journal} {Kristall und Technik}\ }\textbf
  {\bibinfo {volume} {16}},\ \bibinfo {pages} {209} (\bibinfo {year}
  {1981})}\BibitemShut {NoStop}%
\bibitem [{\citenamefont {Watkins}\ and\ \citenamefont
  {Pantelides}(1986)}]{EPR2}%
  \BibitemOpen
  \bibfield  {author} {\bibinfo {author} {\bibfnamefont {G.}~\bibnamefont
  {Watkins}}\ and\ \bibinfo {author} {\bibfnamefont {S.}~\bibnamefont
  {Pantelides}},\ }\href@noop {} {\bibfield  {journal} {\bibinfo  {journal}
  {Gordon Breach, New York}\ } (\bibinfo {year} {1986})}\BibitemShut {NoStop}%
\bibitem [{\citenamefont {Watkins}(2000)}]{Last40Years}%
  \BibitemOpen
  \bibfield  {author} {\bibinfo {author} {\bibfnamefont {G.~D.}\ \bibnamefont
  {Watkins}},\ }\href {\doibase https://doi.org/10.1016/S1369-8001(00)00037-8}
  {\bibfield  {journal} {\bibinfo  {journal} {Materials Science in
  Semiconductor Processing}\ }\textbf {\bibinfo {volume} {3}},\ \bibinfo
  {pages} {227} (\bibinfo {year} {2000})}\BibitemShut {NoStop}%
\bibitem [{\citenamefont {Breitenstein}\ and\ \citenamefont
  {Heydenreich}(1985)}]{DLTS}%
  \BibitemOpen
  \bibfield  {author} {\bibinfo {author} {\bibfnamefont {O.}~\bibnamefont
  {Breitenstein}}\ and\ \bibinfo {author} {\bibfnamefont {J.}~\bibnamefont
  {Heydenreich}},\ }\href {\doibase https://doi.org/10.1002/sca.4950070602}
  {\bibfield  {journal} {\bibinfo  {journal} {Scanning}\ }\textbf {\bibinfo
  {volume} {7}},\ \bibinfo {pages} {273} (\bibinfo {year} {1985})}\BibitemShut
  {NoStop}%
\bibitem [{\citenamefont {Fukata}\ and\ \citenamefont {Suezawa}(1999)}]{OAS}%
  \BibitemOpen
  \bibfield  {author} {\bibinfo {author} {\bibfnamefont {N.}~\bibnamefont
  {Fukata}}\ and\ \bibinfo {author} {\bibfnamefont {M.}~\bibnamefont
  {Suezawa}},\ }\href {\doibase 10.1063/1.370978} {\bibfield  {journal}
  {\bibinfo  {journal} {Journal of Applied Physics}\ }\textbf {\bibinfo
  {volume} {86}},\ \bibinfo {pages} {1848} (\bibinfo {year}
  {1999})}\BibitemShut {NoStop}%
\bibitem [{\citenamefont {Tilley}(2018)}]{ThermodynamicStability}%
  \BibitemOpen
  \bibfield  {author} {\bibinfo {author} {\bibfnamefont {R.~J.}\ \bibnamefont
  {Tilley}},\ }in\ \href {\doibase
  https://doi.org/10.1002/9781119951438.eibc0058.pub2} {\emph {\bibinfo
  {booktitle} {Encyclopedia of Inorganic and Bioinorganic Chemistry}}}\
  (\bibinfo  {publisher} {John Wiley and Sons, Ltd},\ \bibinfo {year} {2018})\
  pp.\ \bibinfo {pages} {1--23}\BibitemShut {NoStop}%
\bibitem [{\citenamefont {Fahey}\ \emph
  {et~al.}(1989{\natexlab{a}})\citenamefont {Fahey}, \citenamefont {Griffin},\
  and\ \citenamefont {Plummer}}]{Doping1}%
  \BibitemOpen
  \bibfield  {author} {\bibinfo {author} {\bibfnamefont {P.~M.}\ \bibnamefont
  {Fahey}}, \bibinfo {author} {\bibfnamefont {P.~B.}\ \bibnamefont {Griffin}},
  \ and\ \bibinfo {author} {\bibfnamefont {J.~D.}\ \bibnamefont {Plummer}},\
  }\href {\doibase 10.1103/RevModPhys.61.289} {\bibfield  {journal} {\bibinfo
  {journal} {Rev. Mod. Phys.}\ }\textbf {\bibinfo {volume} {61}},\ \bibinfo
  {pages} {289} (\bibinfo {year} {1989}{\natexlab{a}})}\BibitemShut {NoStop}%
\bibitem [{\citenamefont {Eaglesham}\ \emph {et~al.}(1994)\citenamefont
  {Eaglesham}, \citenamefont {Stolk}, \citenamefont {Gossmann},\ and\
  \citenamefont {Poate}}]{Doping2}%
  \BibitemOpen
  \bibfield  {author} {\bibinfo {author} {\bibfnamefont {D.~J.}\ \bibnamefont
  {Eaglesham}}, \bibinfo {author} {\bibfnamefont {P.~A.}\ \bibnamefont
  {Stolk}}, \bibinfo {author} {\bibfnamefont {H.}~\bibnamefont {Gossmann}}, \
  and\ \bibinfo {author} {\bibfnamefont {J.~M.}\ \bibnamefont {Poate}},\ }\href
  {https://doi.org/10.1063/1.112725} {\bibfield  {journal} {\bibinfo  {journal}
  {Applied Physics Letters}\ }\textbf {\bibinfo {volume} {65}},\ \bibinfo
  {pages} {2305} (\bibinfo {year} {1994})}\BibitemShut {NoStop}%
\bibitem [{\citenamefont {Richie}\ \emph {et~al.}(2004)\citenamefont {Richie},
  \citenamefont {Kim}, \citenamefont {Barr}, \citenamefont {Hazzard},
  \citenamefont {Hennig},\ and\ \citenamefont {Wilkins}}]{Doping3}%
  \BibitemOpen
  \bibfield  {author} {\bibinfo {author} {\bibfnamefont {D.~A.}\ \bibnamefont
  {Richie}}, \bibinfo {author} {\bibfnamefont {J.}~\bibnamefont {Kim}},
  \bibinfo {author} {\bibfnamefont {S.~A.}\ \bibnamefont {Barr}}, \bibinfo
  {author} {\bibfnamefont {K.~R.~A.}\ \bibnamefont {Hazzard}}, \bibinfo
  {author} {\bibfnamefont {R.}~\bibnamefont {Hennig}}, \ and\ \bibinfo {author}
  {\bibfnamefont {J.~W.}\ \bibnamefont {Wilkins}},\ }\href {\doibase
  10.1103/PhysRevLett.92.045501} {\bibfield  {journal} {\bibinfo  {journal}
  {Phys. Rev. Lett.}\ }\textbf {\bibinfo {volume} {92}},\ \bibinfo {pages}
  {045501} (\bibinfo {year} {2004})}\BibitemShut {NoStop}%
\bibitem [{\citenamefont {Vaidyanathan}\ \emph
  {et~al.}(2007{\natexlab{a}})\citenamefont {Vaidyanathan}, \citenamefont
  {Jung},\ and\ \citenamefont {Seebauer}}]{Doping4}%
  \BibitemOpen
  \bibfield  {author} {\bibinfo {author} {\bibfnamefont {R.}~\bibnamefont
  {Vaidyanathan}}, \bibinfo {author} {\bibfnamefont {M.~Y.~L.}\ \bibnamefont
  {Jung}}, \ and\ \bibinfo {author} {\bibfnamefont {E.~G.}\ \bibnamefont
  {Seebauer}},\ }\href {\doibase 10.1103/PhysRevB.75.195209} {\bibfield
  {journal} {\bibinfo  {journal} {PHYSICAL REVIEW B}\ }\textbf {\bibinfo
  {volume} {75}} (\bibinfo {year} {2007}{\natexlab{a}}),\
  10.1103/PhysRevB.75.195209}\BibitemShut {NoStop}%
\bibitem [{\citenamefont {Riedel}\ \emph {et~al.}(2012)\citenamefont {Riedel},
  \citenamefont {Fuchs}, \citenamefont {Kraus}, \citenamefont {V\"ath},
  \citenamefont {Sperlich}, \citenamefont {Dyakonov}, \citenamefont
  {Soltamova}, \citenamefont {Baranov}, \citenamefont {Ilyin},\ and\
  \citenamefont {Astakhov}}]{qd1}%
  \BibitemOpen
  \bibfield  {author} {\bibinfo {author} {\bibfnamefont {D.}~\bibnamefont
  {Riedel}}, \bibinfo {author} {\bibfnamefont {F.}~\bibnamefont {Fuchs}},
  \bibinfo {author} {\bibfnamefont {H.}~\bibnamefont {Kraus}}, \bibinfo
  {author} {\bibfnamefont {S.}~\bibnamefont {V\"ath}}, \bibinfo {author}
  {\bibfnamefont {A.}~\bibnamefont {Sperlich}}, \bibinfo {author}
  {\bibfnamefont {V.}~\bibnamefont {Dyakonov}}, \bibinfo {author}
  {\bibfnamefont {A.~A.}\ \bibnamefont {Soltamova}}, \bibinfo {author}
  {\bibfnamefont {P.~G.}\ \bibnamefont {Baranov}}, \bibinfo {author}
  {\bibfnamefont {V.~A.}\ \bibnamefont {Ilyin}}, \ and\ \bibinfo {author}
  {\bibfnamefont {G.~V.}\ \bibnamefont {Astakhov}},\ }\href {\doibase
  10.1103/PhysRevLett.109.226402} {\bibfield  {journal} {\bibinfo  {journal}
  {Phys. Rev. Lett.}\ }\textbf {\bibinfo {volume} {109}},\ \bibinfo {pages}
  {226402} (\bibinfo {year} {2012})}\BibitemShut {NoStop}%
\bibitem [{\citenamefont {Tyryshkin}\ \emph {et~al.}(2011)\citenamefont
  {Tyryshkin}, \citenamefont {Tojo}, \citenamefont {Morton}, \citenamefont
  {Riemann}, \citenamefont {Abrosimov}, \citenamefont {Becker}, \citenamefont
  {Pohl}, \citenamefont {Schenkel}, \citenamefont {Thewalt}, \citenamefont
  {Itoh},\ and\ \citenamefont {Lyon}}]{qd2}%
  \BibitemOpen
  \bibfield  {author} {\bibinfo {author} {\bibfnamefont {A.~M.}\ \bibnamefont
  {Tyryshkin}}, \bibinfo {author} {\bibfnamefont {S.}~\bibnamefont {Tojo}},
  \bibinfo {author} {\bibfnamefont {J.~J.~L.}\ \bibnamefont {Morton}}, \bibinfo
  {author} {\bibfnamefont {H.}~\bibnamefont {Riemann}}, \bibinfo {author}
  {\bibfnamefont {N.~V.}\ \bibnamefont {Abrosimov}}, \bibinfo {author}
  {\bibfnamefont {P.}~\bibnamefont {Becker}}, \bibinfo {author} {\bibfnamefont
  {H.-J.}\ \bibnamefont {Pohl}}, \bibinfo {author} {\bibfnamefont
  {T.}~\bibnamefont {Schenkel}}, \bibinfo {author} {\bibfnamefont {M.~L.~W.}\
  \bibnamefont {Thewalt}}, \bibinfo {author} {\bibfnamefont {K.~M.}\
  \bibnamefont {Itoh}}, \ and\ \bibinfo {author} {\bibfnamefont {S.~A.}\
  \bibnamefont {Lyon}},\ }\href {\doibase 10.1038/nmat3182} {\bibfield
  {journal} {\bibinfo  {journal} {Nature Materials}\ }\textbf {\bibinfo
  {volume} {11}},\ \bibinfo {pages} {143} (\bibinfo {year} {2011})}\BibitemShut
  {NoStop}%
\bibitem [{\citenamefont {Otsuka}\ and\ \citenamefont
  {Ren}(1999)}]{ShapeMemory}%
  \BibitemOpen
  \bibfield  {author} {\bibinfo {author} {\bibfnamefont {K.}~\bibnamefont
  {Otsuka}}\ and\ \bibinfo {author} {\bibfnamefont {X.}~\bibnamefont {Ren}},\
  }\href {\doibase https://doi.org/10.1016/S0966-9795(98)00070-3} {\bibfield
  {journal} {\bibinfo  {journal} {Intermetallics}\ }\textbf {\bibinfo {volume}
  {7}},\ \bibinfo {pages} {511} (\bibinfo {year} {1999})}\BibitemShut {NoStop}%
\bibitem [{\citenamefont {Gao}\ and\ \citenamefont
  {Tkatchenko}(2013)}]{experimentalData}%
  \BibitemOpen
  \bibfield  {author} {\bibinfo {author} {\bibfnamefont {W.}~\bibnamefont
  {Gao}}\ and\ \bibinfo {author} {\bibfnamefont {A.}~\bibnamefont
  {Tkatchenko}},\ }\href {\doibase 10.1103/PhysRevLett.111.045501} {\bibfield
  {journal} {\bibinfo  {journal} {Phys. Rev. Lett.}\ }\textbf {\bibinfo
  {volume} {111}},\ \bibinfo {pages} {045501} (\bibinfo {year}
  {2013})}\BibitemShut {NoStop}%
\bibitem [{\citenamefont {Bruneval}(2012)}]{dow5}%
  \BibitemOpen
  \bibfield  {author} {\bibinfo {author} {\bibfnamefont {F.}~\bibnamefont
  {Bruneval}},\ }\href {\doibase 10.1103/PhysRevLett.108.256403} {\bibfield
  {journal} {\bibinfo  {journal} {Phys. Rev. Lett.}\ }\textbf {\bibinfo
  {volume} {108}},\ \bibinfo {pages} {256403} (\bibinfo {year}
  {2012})}\BibitemShut {NoStop}%
\bibitem [{\citenamefont {Ramprasad}\ \emph {et~al.}(2012)\citenamefont
  {Ramprasad}, \citenamefont {Zhu}, \citenamefont {Rinke},\ and\ \citenamefont
  {Scheffler}}]{dow6}%
  \BibitemOpen
  \bibfield  {author} {\bibinfo {author} {\bibfnamefont {R.}~\bibnamefont
  {Ramprasad}}, \bibinfo {author} {\bibfnamefont {H.}~\bibnamefont {Zhu}},
  \bibinfo {author} {\bibfnamefont {P.}~\bibnamefont {Rinke}}, \ and\ \bibinfo
  {author} {\bibfnamefont {M.}~\bibnamefont {Scheffler}},\ }\href {\doibase
  10.1103/PhysRevLett.108.066404} {\bibfield  {journal} {\bibinfo  {journal}
  {Phys. Rev. Lett.}\ }\textbf {\bibinfo {volume} {108}},\ \bibinfo {pages}
  {066404} (\bibinfo {year} {2012})}\BibitemShut {NoStop}%
\bibitem [{\citenamefont {Rinke}\ \emph {et~al.}(2009)\citenamefont {Rinke},
  \citenamefont {Janotti}, \citenamefont {Scheffler},\ and\ \citenamefont
  {Van~de Walle}}]{dow7}%
  \BibitemOpen
  \bibfield  {author} {\bibinfo {author} {\bibfnamefont {P.}~\bibnamefont
  {Rinke}}, \bibinfo {author} {\bibfnamefont {A.}~\bibnamefont {Janotti}},
  \bibinfo {author} {\bibfnamefont {M.}~\bibnamefont {Scheffler}}, \ and\
  \bibinfo {author} {\bibfnamefont {C.~G.}\ \bibnamefont {Van~de Walle}},\
  }\href {\doibase 10.1103/PhysRevLett.102.026402} {\bibfield  {journal}
  {\bibinfo  {journal} {Phys. Rev. Lett.}\ }\textbf {\bibinfo {volume} {102}},\
  \bibinfo {pages} {026402} (\bibinfo {year} {2009})}\BibitemShut {NoStop}%
\bibitem [{\citenamefont {Van~de Walle}\ and\ \citenamefont
  {Janotti}(2011)}]{dow8}%
  \BibitemOpen
  \bibfield  {author} {\bibinfo {author} {\bibfnamefont {C.~G.}\ \bibnamefont
  {Van~de Walle}}\ and\ \bibinfo {author} {\bibfnamefont {A.}~\bibnamefont
  {Janotti}},\ }\href {\doibase https://doi.org/10.1002/pssb.201046290}
  {\bibfield  {journal} {\bibinfo  {journal} {physica status solidi (b)}\
  }\textbf {\bibinfo {volume} {248}},\ \bibinfo {pages} {19} (\bibinfo {year}
  {2011})}\BibitemShut {NoStop}%
\bibitem [{\citenamefont {Van~de Walle}\ and\ \citenamefont
  {Neugebauer}(2004)}]{dow9}%
  \BibitemOpen
  \bibfield  {author} {\bibinfo {author} {\bibfnamefont {C.~G.}\ \bibnamefont
  {Van~de Walle}}\ and\ \bibinfo {author} {\bibfnamefont {J.}~\bibnamefont
  {Neugebauer}},\ }\href {\doibase 10.1063/1.1682673} {\bibfield  {journal}
  {\bibinfo  {journal} {Journal of Applied Physics}\ }\textbf {\bibinfo
  {volume} {95}},\ \bibinfo {pages} {3851} (\bibinfo {year}
  {2004})}\BibitemShut {NoStop}%
\bibitem [{\citenamefont {Batista}\ \emph {et~al.}(2006)\citenamefont
  {Batista}, \citenamefont {Heyd}, \citenamefont {Hennig}, \citenamefont
  {Uberuaga}, \citenamefont {Martin}, \citenamefont {Scuseria}, \citenamefont
  {Umrigar},\ and\ \citenamefont {Wilkins}}]{dow10}%
  \BibitemOpen
  \bibfield  {author} {\bibinfo {author} {\bibfnamefont {E.~R.}\ \bibnamefont
  {Batista}}, \bibinfo {author} {\bibfnamefont {J.}~\bibnamefont {Heyd}},
  \bibinfo {author} {\bibfnamefont {R.~G.}\ \bibnamefont {Hennig}}, \bibinfo
  {author} {\bibfnamefont {B.~P.}\ \bibnamefont {Uberuaga}}, \bibinfo {author}
  {\bibfnamefont {R.~L.}\ \bibnamefont {Martin}}, \bibinfo {author}
  {\bibfnamefont {G.~E.}\ \bibnamefont {Scuseria}}, \bibinfo {author}
  {\bibfnamefont {C.~J.}\ \bibnamefont {Umrigar}}, \ and\ \bibinfo {author}
  {\bibfnamefont {J.~W.}\ \bibnamefont {Wilkins}},\ }\href {\doibase
  10.1103/PhysRevB.74.121102} {\bibfield  {journal} {\bibinfo  {journal} {Phys.
  Rev. B}\ }\textbf {\bibinfo {volume} {74}},\ \bibinfo {pages} {121102}
  (\bibinfo {year} {2006})}\BibitemShut {NoStop}%
\bibitem [{\citenamefont {Leung}\ \emph {et~al.}(1999)\citenamefont {Leung},
  \citenamefont {Needs}, \citenamefont {Rajagopal}, \citenamefont {Itoh},\ and\
  \citenamefont {Ihara}}]{dow11}%
  \BibitemOpen
  \bibfield  {author} {\bibinfo {author} {\bibfnamefont {W.-K.}\ \bibnamefont
  {Leung}}, \bibinfo {author} {\bibfnamefont {R.~J.}\ \bibnamefont {Needs}},
  \bibinfo {author} {\bibfnamefont {G.}~\bibnamefont {Rajagopal}}, \bibinfo
  {author} {\bibfnamefont {S.}~\bibnamefont {Itoh}}, \ and\ \bibinfo {author}
  {\bibfnamefont {S.}~\bibnamefont {Ihara}},\ }\href {\doibase
  10.1103/PhysRevLett.83.2351} {\bibfield  {journal} {\bibinfo  {journal}
  {Phys. Rev. Lett.}\ }\textbf {\bibinfo {volume} {83}},\ \bibinfo {pages}
  {2351} (\bibinfo {year} {1999})}\BibitemShut {NoStop}%
\bibitem [{\citenamefont {Bl\"ochl}\ \emph {et~al.}(1993)\citenamefont
  {Bl\"ochl}, \citenamefont {Smargiassi}, \citenamefont {Car}, \citenamefont
  {Laks}, \citenamefont {Andreoni},\ and\ \citenamefont {Pantelides}}]{dow12}%
  \BibitemOpen
  \bibfield  {author} {\bibinfo {author} {\bibfnamefont {P.~E.}\ \bibnamefont
  {Bl\"ochl}}, \bibinfo {author} {\bibfnamefont {E.}~\bibnamefont
  {Smargiassi}}, \bibinfo {author} {\bibfnamefont {R.}~\bibnamefont {Car}},
  \bibinfo {author} {\bibfnamefont {D.~B.}\ \bibnamefont {Laks}}, \bibinfo
  {author} {\bibfnamefont {W.}~\bibnamefont {Andreoni}}, \ and\ \bibinfo
  {author} {\bibfnamefont {S.~T.}\ \bibnamefont {Pantelides}},\ }\href
  {\doibase 10.1103/PhysRevLett.70.2435} {\bibfield  {journal} {\bibinfo
  {journal} {Phys. Rev. Lett.}\ }\textbf {\bibinfo {volume} {70}},\ \bibinfo
  {pages} {2435} (\bibinfo {year} {1993})}\BibitemShut {NoStop}%
\bibitem [{\citenamefont {Bar-Yam}\ and\ \citenamefont
  {Joannopoulos}(1984)}]{dow13}%
  \BibitemOpen
  \bibfield  {author} {\bibinfo {author} {\bibfnamefont {Y.}~\bibnamefont
  {Bar-Yam}}\ and\ \bibinfo {author} {\bibfnamefont {J.~D.}\ \bibnamefont
  {Joannopoulos}},\ }\href {\doibase 10.1103/PhysRevB.30.1844} {\bibfield
  {journal} {\bibinfo  {journal} {Phys. Rev. B}\ }\textbf {\bibinfo {volume}
  {30}},\ \bibinfo {pages} {1844} (\bibinfo {year} {1984})}\BibitemShut
  {NoStop}%
\bibitem [{\citenamefont {Vaidyanathan}\ \emph
  {et~al.}(2007{\natexlab{b}})\citenamefont {Vaidyanathan}, \citenamefont
  {Jung},\ and\ \citenamefont {Seebauer}}]{dow14}%
  \BibitemOpen
  \bibfield  {author} {\bibinfo {author} {\bibfnamefont {R.}~\bibnamefont
  {Vaidyanathan}}, \bibinfo {author} {\bibfnamefont {M.~Y.~L.}\ \bibnamefont
  {Jung}}, \ and\ \bibinfo {author} {\bibfnamefont {E.~G.}\ \bibnamefont
  {Seebauer}},\ }\href {\doibase 10.1103/PhysRevB.75.195209} {\bibfield
  {journal} {\bibinfo  {journal} {Phys. Rev. B}\ }\textbf {\bibinfo {volume}
  {75}},\ \bibinfo {pages} {195209} (\bibinfo {year}
  {2007}{\natexlab{b}})}\BibitemShut {NoStop}%
\bibitem [{\citenamefont {Ranki}\ and\ \citenamefont {Saarinen}(2004)}]{dow15}%
  \BibitemOpen
  \bibfield  {author} {\bibinfo {author} {\bibfnamefont {V.}~\bibnamefont
  {Ranki}}\ and\ \bibinfo {author} {\bibfnamefont {K.}~\bibnamefont
  {Saarinen}},\ }\href {\doibase 10.1103/PhysRevLett.93.255502} {\bibfield
  {journal} {\bibinfo  {journal} {Phys. Rev. Lett.}\ }\textbf {\bibinfo
  {volume} {93}},\ \bibinfo {pages} {255502} (\bibinfo {year}
  {2004})}\BibitemShut {NoStop}%
\bibitem [{\citenamefont {Bracht}\ \emph {et~al.}(2003)\citenamefont {Bracht},
  \citenamefont {Pedersen}, \citenamefont {Zangenberg}, \citenamefont {Larsen},
  \citenamefont {Haller}, \citenamefont {Lulli},\ and\ \citenamefont
  {Posselt}}]{dow16}%
  \BibitemOpen
  \bibfield  {author} {\bibinfo {author} {\bibfnamefont {H.}~\bibnamefont
  {Bracht}}, \bibinfo {author} {\bibfnamefont {J.~F.}\ \bibnamefont
  {Pedersen}}, \bibinfo {author} {\bibfnamefont {N.}~\bibnamefont
  {Zangenberg}}, \bibinfo {author} {\bibfnamefont {A.~N.}\ \bibnamefont
  {Larsen}}, \bibinfo {author} {\bibfnamefont {E.~E.}\ \bibnamefont {Haller}},
  \bibinfo {author} {\bibfnamefont {G.}~\bibnamefont {Lulli}}, \ and\ \bibinfo
  {author} {\bibfnamefont {M.}~\bibnamefont {Posselt}},\ }\href {\doibase
  10.1103/PhysRevLett.91.245502} {\bibfield  {journal} {\bibinfo  {journal}
  {Phys. Rev. Lett.}\ }\textbf {\bibinfo {volume} {91}},\ \bibinfo {pages}
  {245502} (\bibinfo {year} {2003})}\BibitemShut {NoStop}%
\bibitem [{\citenamefont {Ural}\ \emph
  {et~al.}(1999{\natexlab{a}})\citenamefont {Ural}, \citenamefont {Griffin},\
  and\ \citenamefont {Plummer}}]{dow17}%
  \BibitemOpen
  \bibfield  {author} {\bibinfo {author} {\bibfnamefont {A.}~\bibnamefont
  {Ural}}, \bibinfo {author} {\bibfnamefont {P.~B.}\ \bibnamefont {Griffin}}, \
  and\ \bibinfo {author} {\bibfnamefont {J.~D.}\ \bibnamefont {Plummer}},\
  }\href {\doibase 10.1103/PhysRevLett.83.3454} {\bibfield  {journal} {\bibinfo
   {journal} {Phys. Rev. Lett.}\ }\textbf {\bibinfo {volume} {83}},\ \bibinfo
  {pages} {3454} (\bibinfo {year} {1999}{\natexlab{a}})}\BibitemShut {NoStop}%
\bibitem [{\citenamefont {Bracht}\ \emph
  {et~al.}(1998{\natexlab{a}})\citenamefont {Bracht}, \citenamefont {Haller},\
  and\ \citenamefont {Clark-Phelps}}]{dow18}%
  \BibitemOpen
  \bibfield  {author} {\bibinfo {author} {\bibfnamefont {H.}~\bibnamefont
  {Bracht}}, \bibinfo {author} {\bibfnamefont {E.~E.}\ \bibnamefont {Haller}},
  \ and\ \bibinfo {author} {\bibfnamefont {R.}~\bibnamefont {Clark-Phelps}},\
  }\href {\doibase 10.1103/PhysRevLett.81.393} {\bibfield  {journal} {\bibinfo
  {journal} {Phys. Rev. Lett.}\ }\textbf {\bibinfo {volume} {81}},\ \bibinfo
  {pages} {393} (\bibinfo {year} {1998}{\natexlab{a}})}\BibitemShut {NoStop}%
\bibitem [{\citenamefont {Bracht}\ \emph
  {et~al.}(1995{\natexlab{a}})\citenamefont {Bracht}, \citenamefont
  {Stolwijk},\ and\ \citenamefont {Mehrer}}]{dow19}%
  \BibitemOpen
  \bibfield  {author} {\bibinfo {author} {\bibfnamefont {H.}~\bibnamefont
  {Bracht}}, \bibinfo {author} {\bibfnamefont {N.~A.}\ \bibnamefont
  {Stolwijk}}, \ and\ \bibinfo {author} {\bibfnamefont {H.}~\bibnamefont
  {Mehrer}},\ }\href {\doibase 10.1103/PhysRevB.52.16542} {\bibfield  {journal}
  {\bibinfo  {journal} {Phys. Rev. B}\ }\textbf {\bibinfo {volume} {52}},\
  \bibinfo {pages} {16542} (\bibinfo {year} {1995}{\natexlab{a}})}\BibitemShut
  {NoStop}%
\bibitem [{\citenamefont {Bracht}\ \emph {et~al.}(2007)\citenamefont {Bracht},
  \citenamefont {Silvestri}, \citenamefont {Sharp},\ and\ \citenamefont
  {Haller}}]{dow20}%
  \BibitemOpen
  \bibfield  {author} {\bibinfo {author} {\bibfnamefont {H.}~\bibnamefont
  {Bracht}}, \bibinfo {author} {\bibfnamefont {H.~H.}\ \bibnamefont
  {Silvestri}}, \bibinfo {author} {\bibfnamefont {I.~D.}\ \bibnamefont
  {Sharp}}, \ and\ \bibinfo {author} {\bibfnamefont {E.~E.}\ \bibnamefont
  {Haller}},\ }\href {\doibase 10.1103/PhysRevB.75.035211} {\bibfield
  {journal} {\bibinfo  {journal} {Phys. Rev. B}\ }\textbf {\bibinfo {volume}
  {75}},\ \bibinfo {pages} {035211} (\bibinfo {year} {2007})}\BibitemShut
  {NoStop}%
\bibitem [{\citenamefont {Shimizu}\ \emph {et~al.}(2007)\citenamefont
  {Shimizu}, \citenamefont {Uematsu},\ and\ \citenamefont {Itoh}}]{dow21}%
  \BibitemOpen
  \bibfield  {author} {\bibinfo {author} {\bibfnamefont {Y.}~\bibnamefont
  {Shimizu}}, \bibinfo {author} {\bibfnamefont {M.}~\bibnamefont {Uematsu}}, \
  and\ \bibinfo {author} {\bibfnamefont {K.~M.}\ \bibnamefont {Itoh}},\ }\href
  {\doibase 10.1103/PhysRevLett.98.095901} {\bibfield  {journal} {\bibinfo
  {journal} {Phys. Rev. Lett.}\ }\textbf {\bibinfo {volume} {98}},\ \bibinfo
  {pages} {095901} (\bibinfo {year} {2007})}\BibitemShut {NoStop}%
\bibitem [{\citenamefont {Fahey}\ \emph
  {et~al.}(1989{\natexlab{b}})\citenamefont {Fahey}, \citenamefont {Griffin},\
  and\ \citenamefont {Plummer}}]{dow22}%
  \BibitemOpen
  \bibfield  {author} {\bibinfo {author} {\bibfnamefont {P.~M.}\ \bibnamefont
  {Fahey}}, \bibinfo {author} {\bibfnamefont {P.~B.}\ \bibnamefont {Griffin}},
  \ and\ \bibinfo {author} {\bibfnamefont {J.~D.}\ \bibnamefont {Plummer}},\
  }\href {\doibase 10.1103/RevModPhys.61.289} {\bibfield  {journal} {\bibinfo
  {journal} {Rev. Mod. Phys.}\ }\textbf {\bibinfo {volume} {61}},\ \bibinfo
  {pages} {289} (\bibinfo {year} {1989}{\natexlab{b}})}\BibitemShut {NoStop}%
\bibitem [{\citenamefont {Dannefaer}\ \emph {et~al.}(1986)\citenamefont
  {Dannefaer}, \citenamefont {Mascher},\ and\ \citenamefont {Kerr}}]{dow23}%
  \BibitemOpen
  \bibfield  {author} {\bibinfo {author} {\bibfnamefont {S.}~\bibnamefont
  {Dannefaer}}, \bibinfo {author} {\bibfnamefont {P.}~\bibnamefont {Mascher}},
  \ and\ \bibinfo {author} {\bibfnamefont {D.}~\bibnamefont {Kerr}},\ }\href
  {\doibase 10.1103/PhysRevLett.56.2195} {\bibfield  {journal} {\bibinfo
  {journal} {Phys. Rev. Lett.}\ }\textbf {\bibinfo {volume} {56}},\ \bibinfo
  {pages} {2195} (\bibinfo {year} {1986})}\BibitemShut {NoStop}%
\bibitem [{\citenamefont {Ural}\ \emph
  {et~al.}(1999{\natexlab{b}})\citenamefont {Ural}, \citenamefont {Griffin},\
  and\ \citenamefont {Plummer}}]{exp6}%
  \BibitemOpen
  \bibfield  {author} {\bibinfo {author} {\bibfnamefont {A.}~\bibnamefont
  {Ural}}, \bibinfo {author} {\bibfnamefont {P.~B.}\ \bibnamefont {Griffin}}, \
  and\ \bibinfo {author} {\bibfnamefont {J.~D.}\ \bibnamefont {Plummer}},\
  }\href {\doibase 10.1063/1.370285} {\bibfield  {journal} {\bibinfo  {journal}
  {Journal of Applied Physics}\ }\textbf {\bibinfo {volume} {85}},\ \bibinfo
  {pages} {6440} (\bibinfo {year} {1999}{\natexlab{b}})}\BibitemShut {NoStop}%
\bibitem [{\citenamefont {Parker}\ \emph {et~al.}(2011)\citenamefont {Parker},
  \citenamefont {Wilkins},\ and\ \citenamefont {Hennig}}]{QMCRef}%
  \BibitemOpen
  \bibfield  {author} {\bibinfo {author} {\bibfnamefont {W.~D.}\ \bibnamefont
  {Parker}}, \bibinfo {author} {\bibfnamefont {J.~W.}\ \bibnamefont {Wilkins}},
  \ and\ \bibinfo {author} {\bibfnamefont {R.~G.}\ \bibnamefont {Hennig}},\
  }\href {\doibase https://doi.org/10.1002/pssb.201046149} {\bibfield
  {journal} {\bibinfo  {journal} {physica status solidi (b)}\ }\textbf
  {\bibinfo {volume} {248}},\ \bibinfo {pages} {267} (\bibinfo {year}
  {2011})}\BibitemShut {NoStop}%
\bibitem [{\citenamefont {Kaltak}\ \emph {et~al.}(2014)\citenamefont {Kaltak},
  \citenamefont {Klime\ifmmode~\check{s}\else \v{s}\fi{}},\ and\ \citenamefont
  {Kresse}}]{Kresse}%
  \BibitemOpen
  \bibfield  {author} {\bibinfo {author} {\bibfnamefont {M.}~\bibnamefont
  {Kaltak}}, \bibinfo {author} {\bibfnamefont {J.~c.~v.}\ \bibnamefont
  {Klime\ifmmode~\check{s}\else \v{s}\fi{}}}, \ and\ \bibinfo {author}
  {\bibfnamefont {G.}~\bibnamefont {Kresse}},\ }\href {\doibase
  10.1103/PhysRevB.90.054115} {\bibfield  {journal} {\bibinfo  {journal} {Phys.
  Rev. B}\ }\textbf {\bibinfo {volume} {90}},\ \bibinfo {pages} {054115}
  (\bibinfo {year} {2014})}\BibitemShut {NoStop}%
\bibitem [{\citenamefont {Leung}\ \emph {et~al.}(2001)\citenamefont {Leung},
  \citenamefont {Needs}, \citenamefont {Rajagopal}, \citenamefont {Itoh},\ and\
  \citenamefont {Ihara}}]{QMCOld}%
  \BibitemOpen
  \bibfield  {author} {\bibinfo {author} {\bibfnamefont {W.~K.}\ \bibnamefont
  {Leung}}, \bibinfo {author} {\bibfnamefont {R.}~\bibnamefont {Needs}},
  \bibinfo {author} {\bibfnamefont {G.}~\bibnamefont {Rajagopal}}, \bibinfo
  {author} {\bibfnamefont {S.}~\bibnamefont {Itoh}}, \ and\ \bibinfo {author}
  {\bibfnamefont {S.}~\bibnamefont {Ihara}},\ }\href {\doibase
  10.1155/2001/83797} {\bibfield  {journal} {\bibinfo  {journal} {VLSI Design}\
  }\textbf {\bibinfo {volume} {13}} (\bibinfo {year} {2001}),\
  10.1155/2001/83797}\BibitemShut {NoStop}%
\bibitem [{\citenamefont {Booth}\ \emph {et~al.}(2013)\citenamefont {Booth},
  \citenamefont {Gr{\"{u}}neis}, \citenamefont {Kresse},\ and\ \citenamefont
  {Alavi}}]{Booth2013}%
  \BibitemOpen
  \bibfield  {author} {\bibinfo {author} {\bibfnamefont {G.~H.}\ \bibnamefont
  {Booth}}, \bibinfo {author} {\bibfnamefont {A.}~\bibnamefont
  {Gr{\"{u}}neis}}, \bibinfo {author} {\bibfnamefont {G.}~\bibnamefont
  {Kresse}}, \ and\ \bibinfo {author} {\bibfnamefont {A.}~\bibnamefont
  {Alavi}},\ }\href {\doibase 10.1038/nature11770} {\bibfield  {journal}
  {\bibinfo  {journal} {Nature}\ }\textbf {\bibinfo {volume} {493}},\ \bibinfo
  {pages} {365} (\bibinfo {year} {2013})}\BibitemShut {NoStop}%
\bibitem [{\citenamefont {Gruber}\ and\ \citenamefont
  {Gr\"uneis}(2018)}]{gruber18b}%
  \BibitemOpen
  \bibfield  {author} {\bibinfo {author} {\bibfnamefont {T.}~\bibnamefont
  {Gruber}}\ and\ \bibinfo {author} {\bibfnamefont {A.}~\bibnamefont
  {Gr\"uneis}},\ }\href {\doibase 10.1103/PhysRevB.98.134108} {\bibfield
  {journal} {\bibinfo  {journal} {Phys. Rev. B}\ }\textbf {\bibinfo {volume}
  {98}},\ \bibinfo {pages} {134108} (\bibinfo {year} {2018})}\BibitemShut
  {NoStop}%
\bibitem [{\citenamefont {Gruber}\ \emph {et~al.}(2018)\citenamefont {Gruber},
  \citenamefont {Liao}, \citenamefont {Tsatsoulis}, \citenamefont {Hummel},\
  and\ \citenamefont {Gr\"uneis}}]{Gruber2018}%
  \BibitemOpen
  \bibfield  {author} {\bibinfo {author} {\bibfnamefont {T.}~\bibnamefont
  {Gruber}}, \bibinfo {author} {\bibfnamefont {K.}~\bibnamefont {Liao}},
  \bibinfo {author} {\bibfnamefont {T.}~\bibnamefont {Tsatsoulis}}, \bibinfo
  {author} {\bibfnamefont {F.}~\bibnamefont {Hummel}}, \ and\ \bibinfo {author}
  {\bibfnamefont {A.}~\bibnamefont {Gr\"uneis}},\ }\href {\doibase
  10.1103/PhysRevX.8.021043} {\bibfield  {journal} {\bibinfo  {journal} {Phys.
  Rev. X}\ }\textbf {\bibinfo {volume} {8}},\ \bibinfo {pages} {021043}
  (\bibinfo {year} {2018})}\BibitemShut {NoStop}%
\bibitem [{\citenamefont {McClain}\ \emph {et~al.}(2017)\citenamefont
  {McClain}, \citenamefont {Sun}, \citenamefont {Chan},\ and\ \citenamefont
  {Berkelbach}}]{mcclain_2017}%
  \BibitemOpen
  \bibfield  {author} {\bibinfo {author} {\bibfnamefont {J.}~\bibnamefont
  {McClain}}, \bibinfo {author} {\bibfnamefont {Q.}~\bibnamefont {Sun}},
  \bibinfo {author} {\bibfnamefont {G.~K.-L.}\ \bibnamefont {Chan}}, \ and\
  \bibinfo {author} {\bibfnamefont {T.~C.}\ \bibnamefont {Berkelbach}},\ }\href
  {\doibase 10.1021/acs.jctc.7b00049} {\bibfield  {journal} {\bibinfo
  {journal} {J. Chem. Theory Comput.}\ }\textbf {\bibinfo {volume} {13}},\
  \bibinfo {pages} {1209} (\bibinfo {year} {2017})},\ \bibinfo {note} {pMID:
  28218843}\BibitemShut {NoStop}%
\bibitem [{\citenamefont {Sch{\"{u}}tz}\ \emph {et~al.}(2017)\citenamefont
  {Sch{\"{u}}tz}, \citenamefont {Maschio}, \citenamefont {Karttunen},\ and\
  \citenamefont {Usvyat}}]{Schutz2017}%
  \BibitemOpen
  \bibfield  {author} {\bibinfo {author} {\bibfnamefont {M.}~\bibnamefont
  {Sch{\"{u}}tz}}, \bibinfo {author} {\bibfnamefont {L.}~\bibnamefont
  {Maschio}}, \bibinfo {author} {\bibfnamefont {A.~J.}\ \bibnamefont
  {Karttunen}}, \ and\ \bibinfo {author} {\bibfnamefont {D.}~\bibnamefont
  {Usvyat}},\ }\href {\doibase 10.1021/acs.jpclett.7b00253} {\bibfield
  {journal} {\bibinfo  {journal} {The Journal of Physical Chemistry Letters}\
  }\textbf {\bibinfo {volume} {8}},\ \bibinfo {pages} {1290} (\bibinfo {year}
  {2017})}\BibitemShut {NoStop}%
\bibitem [{\citenamefont {Usvyat}\ \emph {et~al.}(2018)\citenamefont {Usvyat},
  \citenamefont {Maschio},\ and\ \citenamefont {Sch{\"{u}}tz}}]{Usvyat2018}%
  \BibitemOpen
  \bibfield  {author} {\bibinfo {author} {\bibfnamefont {D.}~\bibnamefont
  {Usvyat}}, \bibinfo {author} {\bibfnamefont {L.}~\bibnamefont {Maschio}}, \
  and\ \bibinfo {author} {\bibfnamefont {M.}~\bibnamefont {Sch{\"{u}}tz}},\
  }\href {\doibase 10.1002/wcms.1357} {\bibfield  {journal} {\bibinfo
  {journal} {WIREs Computational Molecular Science}\ }\textbf {\bibinfo
  {volume} {8}},\ \bibinfo {pages} {1} (\bibinfo {year} {2018})}\BibitemShut
  {NoStop}%
\bibitem [{\citenamefont {Nusspickel}\ and\ \citenamefont
  {Booth}(2022)}]{Nusspickel2021}%
  \BibitemOpen
  \bibfield  {author} {\bibinfo {author} {\bibfnamefont {M.}~\bibnamefont
  {Nusspickel}}\ and\ \bibinfo {author} {\bibfnamefont {G.~H.}\ \bibnamefont
  {Booth}},\ }\href {\doibase 10.1103/PhysRevX.12.011046} {\bibfield  {journal}
  {\bibinfo  {journal} {Phys. Rev. X}\ }\textbf {\bibinfo {volume} {12}},\
  \bibinfo {pages} {011046} (\bibinfo {year} {2022})}\BibitemShut {NoStop}%
\bibitem [{\citenamefont {Chen}\ \emph {et~al.}(2020)\citenamefont {Chen},
  \citenamefont {Bogdanov}, \citenamefont {Usvyat}, \citenamefont {Fang},
  \citenamefont {Michaelides},\ and\ \citenamefont {Alavi}}]{Chen2020}%
  \BibitemOpen
  \bibfield  {author} {\bibinfo {author} {\bibfnamefont {J.}~\bibnamefont
  {Chen}}, \bibinfo {author} {\bibfnamefont {N.~A.}\ \bibnamefont {Bogdanov}},
  \bibinfo {author} {\bibfnamefont {D.}~\bibnamefont {Usvyat}}, \bibinfo
  {author} {\bibfnamefont {W.}~\bibnamefont {Fang}}, \bibinfo {author}
  {\bibfnamefont {A.}~\bibnamefont {Michaelides}}, \ and\ \bibinfo {author}
  {\bibfnamefont {A.}~\bibnamefont {Alavi}},\ }\href {\doibase
  10.1063/5.0030658} {\bibfield  {journal} {\bibinfo  {journal} {The Journal of
  Chemical Physics}\ }\textbf {\bibinfo {volume} {153}},\ \bibinfo {pages}
  {204704} (\bibinfo {year} {2020})}\BibitemShut {NoStop}%
\bibitem [{\citenamefont {Sauer}(2019)}]{Sauer2019}%
  \BibitemOpen
  \bibfield  {author} {\bibinfo {author} {\bibfnamefont {J.}~\bibnamefont
  {Sauer}},\ }\href {\doibase 10.1021/acs.accounts.9b00506} {\bibfield
  {journal} {\bibinfo  {journal} {Accounts of Chemical Research}\ }\textbf
  {\bibinfo {volume} {52}},\ \bibinfo {pages} {3502} (\bibinfo {year}
  {2019})}\BibitemShut {NoStop}%
\bibitem [{\citenamefont {Lin}\ \emph {et~al.}(2020)\citenamefont {Lin},
  \citenamefont {Maschio}, \citenamefont {Kats}, \citenamefont {Usvyat},\ and\
  \citenamefont {Heine}}]{Lin2020}%
  \BibitemOpen
  \bibfield  {author} {\bibinfo {author} {\bibfnamefont {H.~H.}\ \bibnamefont
  {Lin}}, \bibinfo {author} {\bibfnamefont {L.}~\bibnamefont {Maschio}},
  \bibinfo {author} {\bibfnamefont {D.}~\bibnamefont {Kats}}, \bibinfo {author}
  {\bibfnamefont {D.}~\bibnamefont {Usvyat}}, \ and\ \bibinfo {author}
  {\bibfnamefont {T.}~\bibnamefont {Heine}},\ }\href {\doibase
  10.1021/acs.jctc.0c00576} {\bibfield  {journal} {\bibinfo  {journal} {Journal
  of Chemical Theory and Computation}\ }\textbf {\bibinfo {volume} {16}},\
  \bibinfo {pages} {7100} (\bibinfo {year} {2020})}\BibitemShut {NoStop}%
\bibitem [{\citenamefont {Goodpaster}\ \emph {et~al.}(2014)\citenamefont
  {Goodpaster}, \citenamefont {Barnes}, \citenamefont {Manby},\ and\
  \citenamefont {Miller}}]{Goodpaster2014}%
  \BibitemOpen
  \bibfield  {author} {\bibinfo {author} {\bibfnamefont {J.~D.}\ \bibnamefont
  {Goodpaster}}, \bibinfo {author} {\bibfnamefont {T.~A.}\ \bibnamefont
  {Barnes}}, \bibinfo {author} {\bibfnamefont {F.~R.}\ \bibnamefont {Manby}}, \
  and\ \bibinfo {author} {\bibfnamefont {T.~F.}\ \bibnamefont {Miller}},\
  }\href {\doibase 10.1063/1.4864040} {\bibfield  {journal} {\bibinfo
  {journal} {The Journal of Chemical Physics}\ }\textbf {\bibinfo {volume}
  {140}},\ \bibinfo {pages} {18A507} (\bibinfo {year} {2014})}\BibitemShut
  {NoStop}%
\bibitem [{\citenamefont {Sch{\"a}fer}\ \emph {et~al.}(2021)\citenamefont
  {Sch{\"a}fer}, \citenamefont {Libisch}, \citenamefont {Kresse},\ and\
  \citenamefont {Gr{\"u}neis}}]{Schaefer2021a}%
  \BibitemOpen
  \bibfield  {author} {\bibinfo {author} {\bibfnamefont {T.}~\bibnamefont
  {Sch{\"a}fer}}, \bibinfo {author} {\bibfnamefont {F.}~\bibnamefont
  {Libisch}}, \bibinfo {author} {\bibfnamefont {G.}~\bibnamefont {Kresse}}, \
  and\ \bibinfo {author} {\bibfnamefont {A.}~\bibnamefont {Gr{\"u}neis}},\
  }\href {\doibase 10.1063/5.0036363} {\bibfield  {journal} {\bibinfo
  {journal} {The Journal of Chemical Physics}\ }\textbf {\bibinfo {volume}
  {154}},\ \bibinfo {pages} {011101} (\bibinfo {year} {2021})}\BibitemShut
  {NoStop}%
\bibitem [{\citenamefont {Lau}\ \emph {et~al.}(2021)\citenamefont {Lau},
  \citenamefont {Knizia},\ and\ \citenamefont {Berkelbach}}]{Lau2021}%
  \BibitemOpen
  \bibfield  {author} {\bibinfo {author} {\bibfnamefont {B.~T.~G.}\
  \bibnamefont {Lau}}, \bibinfo {author} {\bibfnamefont {G.}~\bibnamefont
  {Knizia}}, \ and\ \bibinfo {author} {\bibfnamefont {T.~C.}\ \bibnamefont
  {Berkelbach}},\ }\href {\doibase 10.1021/acs.jpclett.0c03274} {\bibfield
  {journal} {\bibinfo  {journal} {The Journal of Physical Chemistry Letters}\
  }\textbf {\bibinfo {volume} {12}},\ \bibinfo {pages} {1104} (\bibinfo {year}
  {2021})},\ \bibinfo {note} {pMID: 33475362}\BibitemShut {NoStop}%
\bibitem [{\citenamefont {Fahey}\ \emph
  {et~al.}(1989{\natexlab{c}})\citenamefont {Fahey}, \citenamefont {Griffin},\
  and\ \citenamefont {Plummer}}]{exp2}%
  \BibitemOpen
  \bibfield  {author} {\bibinfo {author} {\bibfnamefont {P.~M.}\ \bibnamefont
  {Fahey}}, \bibinfo {author} {\bibfnamefont {P.~B.}\ \bibnamefont {Griffin}},
  \ and\ \bibinfo {author} {\bibfnamefont {J.~D.}\ \bibnamefont {Plummer}},\
  }\href {\doibase 10.1103/RevModPhys.61.289} {\bibfield  {journal} {\bibinfo
  {journal} {Rev. Mod. Phys.}\ }\textbf {\bibinfo {volume} {61}},\ \bibinfo
  {pages} {289} (\bibinfo {year} {1989}{\natexlab{c}})}\BibitemShut {NoStop}%
\bibitem [{\citenamefont {Bracht}\ \emph
  {et~al.}(1995{\natexlab{b}})\citenamefont {Bracht}, \citenamefont
  {Stolwijk},\ and\ \citenamefont {Mehrer}}]{exp4}%
  \BibitemOpen
  \bibfield  {author} {\bibinfo {author} {\bibfnamefont {H.}~\bibnamefont
  {Bracht}}, \bibinfo {author} {\bibfnamefont {N.~A.}\ \bibnamefont
  {Stolwijk}}, \ and\ \bibinfo {author} {\bibfnamefont {H.}~\bibnamefont
  {Mehrer}},\ }\href {\doibase 10.1103/PhysRevB.52.16542} {\bibfield  {journal}
  {\bibinfo  {journal} {Phys. Rev. B}\ }\textbf {\bibinfo {volume} {52}},\
  \bibinfo {pages} {16542} (\bibinfo {year} {1995}{\natexlab{b}})}\BibitemShut
  {NoStop}%
\bibitem [{\citenamefont {Bracht}\ \emph
  {et~al.}(1998{\natexlab{b}})\citenamefont {Bracht}, \citenamefont {Haller},\
  and\ \citenamefont {Clark-Phelps}}]{exp5}%
  \BibitemOpen
  \bibfield  {author} {\bibinfo {author} {\bibfnamefont {H.}~\bibnamefont
  {Bracht}}, \bibinfo {author} {\bibfnamefont {E.~E.}\ \bibnamefont {Haller}},
  \ and\ \bibinfo {author} {\bibfnamefont {R.}~\bibnamefont {Clark-Phelps}},\
  }\href {\doibase 10.1103/PhysRevLett.81.393} {\bibfield  {journal} {\bibinfo
  {journal} {Phys. Rev. Lett.}\ }\textbf {\bibinfo {volume} {81}},\ \bibinfo
  {pages} {393} (\bibinfo {year} {1998}{\natexlab{b}})}\BibitemShut {NoStop}%
\bibitem [{\citenamefont {Ural}\ \emph
  {et~al.}(1999{\natexlab{c}})\citenamefont {Ural}, \citenamefont {Griffin},\
  and\ \citenamefont {Plummer}}]{exp7}%
  \BibitemOpen
  \bibfield  {author} {\bibinfo {author} {\bibfnamefont {A.}~\bibnamefont
  {Ural}}, \bibinfo {author} {\bibfnamefont {P.~B.}\ \bibnamefont {Griffin}}, \
  and\ \bibinfo {author} {\bibfnamefont {J.~D.}\ \bibnamefont {Plummer}},\
  }\href {\doibase 10.1103/PhysRevLett.83.3454} {\bibfield  {journal} {\bibinfo
   {journal} {Phys. Rev. Lett.}\ }\textbf {\bibinfo {volume} {83}},\ \bibinfo
  {pages} {3454} (\bibinfo {year} {1999}{\natexlab{c}})}\BibitemShut {NoStop}%
\bibitem [{\citenamefont {Raghavachari}\ \emph {et~al.}(1989)\citenamefont
  {Raghavachari}, \citenamefont {Trucks}, \citenamefont {Pople},\ and\
  \citenamefont {Head-Gordon}}]{A.fifth.order.pRaghav1989}%
  \BibitemOpen
  \bibfield  {author} {\bibinfo {author} {\bibfnamefont {K.}~\bibnamefont
  {Raghavachari}}, \bibinfo {author} {\bibfnamefont {G.~W.}\ \bibnamefont
  {Trucks}}, \bibinfo {author} {\bibfnamefont {J.~A.}\ \bibnamefont {Pople}}, \
  and\ \bibinfo {author} {\bibfnamefont {M.}~\bibnamefont {Head-Gordon}},\
  }\href {\doibase 10.1016/s0009-2614(89)87395-6} {\bibfield  {journal}
  {\bibinfo  {journal} {Chemical Physics Letters}\ }\textbf {\bibinfo {volume}
  {157}},\ \bibinfo {pages} {479} (\bibinfo {year} {1989})}\BibitemShut
  {NoStop}%
\bibitem [{\citenamefont {Bartlett}\ and\ \citenamefont
  {Musia\l{}}(2007)}]{bartlett}%
  \BibitemOpen
  \bibfield  {author} {\bibinfo {author} {\bibfnamefont {R.~J.}\ \bibnamefont
  {Bartlett}}\ and\ \bibinfo {author} {\bibfnamefont {M.}~\bibnamefont
  {Musia\l{}}},\ }\href {\doibase 10.1103/RevModPhys.79.291} {\bibfield
  {journal} {\bibinfo  {journal} {Rev. Mod. Phys.}\ }\textbf {\bibinfo {volume}
  {79}},\ \bibinfo {pages} {291} (\bibinfo {year} {2007})}\BibitemShut
  {NoStop}%
\bibitem [{\citenamefont {Isaiah~Shavitt}(2009)}]{book:293288}%
  \BibitemOpen
  \bibfield  {author} {\bibinfo {author} {\bibfnamefont {R.~J.~B.}\
  \bibnamefont {Isaiah~Shavitt}},\ }\href {\doibase 10.1017/CBO9780511596834}
  {}\bibinfo {edition} {1st}\ ed.,\ Cambridge Molecular Science\ (\bibinfo
  {publisher} {Cambridge University Press},\ \bibinfo {year}
  {2009})\BibitemShut {NoStop}%
\bibitem [{\citenamefont {Kato}(1957)}]{On.the.eigenfunKato.1957}%
  \BibitemOpen
  \bibfield  {author} {\bibinfo {author} {\bibfnamefont {T.}~\bibnamefont
  {Kato}},\ }\href {\doibase 10.1002/cpa.3160100201} {\bibfield  {journal}
  {\bibinfo  {journal} {Communications on Pure and Applied Mathematics}\
  }\textbf {\bibinfo {volume} {10}},\ \bibinfo {pages} {151} (\bibinfo {year}
  {1957})}\BibitemShut {NoStop}%
\bibitem [{\citenamefont {Hättig}\ \emph {et~al.}(2012)\citenamefont
  {Hättig}, \citenamefont {Klopper}, \citenamefont {Köhn},\ and\
  \citenamefont {Tew}}]{G17}%
  \BibitemOpen
  \bibfield  {author} {\bibinfo {author} {\bibfnamefont {C.}~\bibnamefont
  {Hättig}}, \bibinfo {author} {\bibfnamefont {W.}~\bibnamefont {Klopper}},
  \bibinfo {author} {\bibfnamefont {A.}~\bibnamefont {Köhn}}, \ and\ \bibinfo
  {author} {\bibfnamefont {D.~P.}\ \bibnamefont {Tew}},\ }\href {\doibase
  10.1021/cr200168z} {\bibfield  {journal} {\bibinfo  {journal} {Chemical
  Reviews}\ }\textbf {\bibinfo {volume} {112}},\ \bibinfo {pages} {4} (\bibinfo
  {year} {2012})},\ \bibinfo {note} {pMID: 22206503}\BibitemShut {NoStop}%
\bibitem [{\citenamefont {Ten-no}(2012)}]{G18}%
  \BibitemOpen
  \bibfield  {author} {\bibinfo {author} {\bibfnamefont {S.}~\bibnamefont
  {Ten-no}},\ }\href {\doibase 10.1007/s00214-011-1070-1} {\bibfield  {journal}
  {\bibinfo  {journal} {Theoretical Chemistry Accounts}\ }\textbf {\bibinfo
  {volume} {131}},\ \bibinfo {pages} {1070} (\bibinfo {year}
  {2012})}\BibitemShut {NoStop}%
\bibitem [{\citenamefont {Gr\"uneis}(2015)}]{G11}%
  \BibitemOpen
  \bibfield  {author} {\bibinfo {author} {\bibfnamefont {A.}~\bibnamefont
  {Gr\"uneis}},\ }\href {\doibase 10.1103/PhysRevLett.115.066402} {\bibfield
  {journal} {\bibinfo  {journal} {Phys. Rev. Lett.}\ }\textbf {\bibinfo
  {volume} {115}},\ \bibinfo {pages} {066402} (\bibinfo {year}
  {2015})}\BibitemShut {NoStop}%
\bibitem [{\citenamefont {Grüneis}\ \emph {et~al.}(2013)\citenamefont
  {Grüneis}, \citenamefont {Shepherd}, \citenamefont {Alavi}, \citenamefont
  {Tew},\ and\ \citenamefont {Booth}}]{G19}%
  \BibitemOpen
  \bibfield  {author} {\bibinfo {author} {\bibfnamefont {A.}~\bibnamefont
  {Grüneis}}, \bibinfo {author} {\bibfnamefont {J.~J.}\ \bibnamefont
  {Shepherd}}, \bibinfo {author} {\bibfnamefont {A.}~\bibnamefont {Alavi}},
  \bibinfo {author} {\bibfnamefont {D.~P.}\ \bibnamefont {Tew}}, \ and\
  \bibinfo {author} {\bibfnamefont {G.~H.}\ \bibnamefont {Booth}},\ }\href
  {\doibase 10.1063/1.4818753} {\bibfield  {journal} {\bibinfo  {journal} {The
  Journal of Chemical Physics}\ }\textbf {\bibinfo {volume} {139}},\ \bibinfo
  {pages} {084112} (\bibinfo {year} {2013})}\BibitemShut {NoStop}%
\bibitem [{\citenamefont {Usvyat}(2013)}]{G20}%
  \BibitemOpen
  \bibfield  {author} {\bibinfo {author} {\bibfnamefont {D.}~\bibnamefont
  {Usvyat}},\ }\href {\doibase 10.1063/1.4829898} {\bibfield  {journal}
  {\bibinfo  {journal} {The Journal of Chemical Physics}\ }\textbf {\bibinfo
  {volume} {139}},\ \bibinfo {pages} {194101} (\bibinfo {year}
  {2013})}\BibitemShut {NoStop}%
\bibitem [{\citenamefont {Irmler}\ \emph {et~al.}(2021)\citenamefont {Irmler},
  \citenamefont {Gallo},\ and\ \citenamefont {Grüneis}}]{FocalPoint}%
  \BibitemOpen
  \bibfield  {author} {\bibinfo {author} {\bibfnamefont {A.}~\bibnamefont
  {Irmler}}, \bibinfo {author} {\bibfnamefont {A.}~\bibnamefont {Gallo}}, \
  and\ \bibinfo {author} {\bibfnamefont {A.}~\bibnamefont {Grüneis}},\ }\href
  {\doibase 10.1063/5.0050054} {\bibfield  {journal} {\bibinfo  {journal} {The
  Journal of Chemical Physics}\ }\textbf {\bibinfo {volume} {154}},\ \bibinfo
  {pages} {234103} (\bibinfo {year} {2021})}\BibitemShut {NoStop}%
\bibitem [{\citenamefont {Liao}\ and\ \citenamefont
  {Grüneis}(2016)}]{GrueneisFiniteSize}%
  \BibitemOpen
  \bibfield  {author} {\bibinfo {author} {\bibfnamefont {K.}~\bibnamefont
  {Liao}}\ and\ \bibinfo {author} {\bibfnamefont {A.}~\bibnamefont
  {Grüneis}},\ }\href {\doibase 10.1063/1.4964307} {\bibfield  {journal}
  {\bibinfo  {journal} {The Journal of Chemical Physics}\ }\textbf {\bibinfo
  {volume} {145}},\ \bibinfo {pages} {141102} (\bibinfo {year}
  {2016})}\BibitemShut {NoStop}%
\bibitem [{\citenamefont {Al-Mushadani}\ and\ \citenamefont
  {Needs}(2003)}]{C3V1}%
  \BibitemOpen
  \bibfield  {author} {\bibinfo {author} {\bibfnamefont {O.~K.}\ \bibnamefont
  {Al-Mushadani}}\ and\ \bibinfo {author} {\bibfnamefont {R.~J.}\ \bibnamefont
  {Needs}},\ }\href {\doibase 10.1103/PhysRevB.68.235205} {\bibfield  {journal}
  {\bibinfo  {journal} {Phys. Rev. B}\ }\textbf {\bibinfo {volume} {68}},\
  \bibinfo {pages} {235205} (\bibinfo {year} {2003})}\BibitemShut {NoStop}%
\bibitem [{\citenamefont {Corsetti}\ and\ \citenamefont
  {Mostofi}(2011)}]{JahnTeller}%
  \BibitemOpen
  \bibfield  {author} {\bibinfo {author} {\bibfnamefont {F.}~\bibnamefont
  {Corsetti}}\ and\ \bibinfo {author} {\bibfnamefont {A.~A.}\ \bibnamefont
  {Mostofi}},\ }\href {\doibase 10.1103/PhysRevB.84.035209} {\bibfield
  {journal} {\bibinfo  {journal} {Phys. Rev. B}\ }\textbf {\bibinfo {volume}
  {84}},\ \bibinfo {pages} {035209} (\bibinfo {year} {2011})}\BibitemShut
  {NoStop}%
\bibitem [{\citenamefont {Perdew}\ \emph {et~al.}(1996)\citenamefont {Perdew},
  \citenamefont {Burke},\ and\ \citenamefont {Ernzerhof}}]{PBE}%
  \BibitemOpen
  \bibfield  {author} {\bibinfo {author} {\bibfnamefont {J.~P.}\ \bibnamefont
  {Perdew}}, \bibinfo {author} {\bibfnamefont {K.}~\bibnamefont {Burke}}, \
  and\ \bibinfo {author} {\bibfnamefont {M.}~\bibnamefont {Ernzerhof}},\ }\href
  {\doibase 10.1103/PhysRevLett.77.3865} {\bibfield  {journal} {\bibinfo
  {journal} {Phys. Rev. Lett.}\ }\textbf {\bibinfo {volume} {77}},\ \bibinfo
  {pages} {3865} (\bibinfo {year} {1996})}\BibitemShut {NoStop}%
\bibitem [{\citenamefont {Kresse}\ and\ \citenamefont {Hafner}(1993)}]{vasp1}%
  \BibitemOpen
  \bibfield  {author} {\bibinfo {author} {\bibfnamefont {G.}~\bibnamefont
  {Kresse}}\ and\ \bibinfo {author} {\bibfnamefont {J.}~\bibnamefont
  {Hafner}},\ }\href {\doibase 10.1103/PhysRevB.47.558} {\bibfield  {journal}
  {\bibinfo  {journal} {Phys. Rev. B}\ }\textbf {\bibinfo {volume} {47}},\
  \bibinfo {pages} {558} (\bibinfo {year} {1993})}\BibitemShut {NoStop}%
\bibitem [{\citenamefont {Kresse}\ and\ \citenamefont
  {Furthm\"uller}(1996)}]{vasp2}%
  \BibitemOpen
  \bibfield  {author} {\bibinfo {author} {\bibfnamefont {G.}~\bibnamefont
  {Kresse}}\ and\ \bibinfo {author} {\bibfnamefont {J.}~\bibnamefont
  {Furthm\"uller}},\ }\href {\doibase 10.1103/PhysRevB.54.11169} {\bibfield
  {journal} {\bibinfo  {journal} {Phys. Rev. B}\ }\textbf {\bibinfo {volume}
  {54}},\ \bibinfo {pages} {11169} (\bibinfo {year} {1996})}\BibitemShut
  {NoStop}%
\bibitem [{\citenamefont {Kresse}\ and\ \citenamefont
  {Furthmüller}(1996)}]{vasp3}%
  \BibitemOpen
  \bibfield  {author} {\bibinfo {author} {\bibfnamefont {G.}~\bibnamefont
  {Kresse}}\ and\ \bibinfo {author} {\bibfnamefont {J.}~\bibnamefont
  {Furthmüller}},\ }\href {\doibase
  https://doi.org/10.1016/0927-0256(96)00008-0} {\bibfield  {journal} {\bibinfo
   {journal} {Computational Materials Science}\ }\textbf {\bibinfo {volume}
  {6}},\ \bibinfo {pages} {15} (\bibinfo {year} {1996})}\BibitemShut {NoStop}%
\bibitem [{\citenamefont {Grüneis}\ \emph {et~al.}(2011)\citenamefont
  {Grüneis}, \citenamefont {Booth}, \citenamefont {Marsman}, \citenamefont
  {Spencer}, \citenamefont {Alavi},\ and\ \citenamefont {Kresse}}]{MP2NO}%
  \BibitemOpen
  \bibfield  {author} {\bibinfo {author} {\bibfnamefont {A.}~\bibnamefont
  {Grüneis}}, \bibinfo {author} {\bibfnamefont {G.~H.}\ \bibnamefont {Booth}},
  \bibinfo {author} {\bibfnamefont {M.}~\bibnamefont {Marsman}}, \bibinfo
  {author} {\bibfnamefont {J.}~\bibnamefont {Spencer}}, \bibinfo {author}
  {\bibfnamefont {A.}~\bibnamefont {Alavi}}, \ and\ \bibinfo {author}
  {\bibfnamefont {G.}~\bibnamefont {Kresse}},\ }\href {\doibase
  10.1021/ct200263g} {\bibfield  {journal} {\bibinfo  {journal} {Journal of
  Chemical Theory and Computation}\ }\textbf {\bibinfo {volume} {7}},\ \bibinfo
  {pages} {2780} (\bibinfo {year} {2011})},\ \bibinfo {note} {pMID:
  26605469}\BibitemShut {NoStop}%
\bibitem [{\citenamefont {Schäfer}\ \emph {et~al.}(2017)\citenamefont
  {Schäfer}, \citenamefont {Ramberger},\ and\ \citenamefont {Kresse}}]{LTMP2}%
  \BibitemOpen
  \bibfield  {author} {\bibinfo {author} {\bibfnamefont {T.}~\bibnamefont
  {Schäfer}}, \bibinfo {author} {\bibfnamefont {B.}~\bibnamefont {Ramberger}},
  \ and\ \bibinfo {author} {\bibfnamefont {G.}~\bibnamefont {Kresse}},\ }\href
  {\doibase 10.1063/1.4976937} {\bibfield  {journal} {\bibinfo  {journal} {The
  Journal of Chemical Physics}\ }\textbf {\bibinfo {volume} {146}},\ \bibinfo
  {pages} {104101} (\bibinfo {year} {2017})}\BibitemShut {NoStop}%
\bibitem [{\citenamefont {Marsman}\ \emph {et~al.}(2009)\citenamefont
  {Marsman}, \citenamefont {Grüneis}, \citenamefont {Paier},\ and\
  \citenamefont {Kresse}}]{Marsman2009}%
  \BibitemOpen
  \bibfield  {author} {\bibinfo {author} {\bibfnamefont {M.}~\bibnamefont
  {Marsman}}, \bibinfo {author} {\bibfnamefont {A.}~\bibnamefont {Grüneis}},
  \bibinfo {author} {\bibfnamefont {J.}~\bibnamefont {Paier}}, \ and\ \bibinfo
  {author} {\bibfnamefont {G.}~\bibnamefont {Kresse}},\ }\href {\doibase
  10.1063/1.3126249} {\bibfield  {journal} {\bibinfo  {journal} {The Journal of
  Chemical Physics}\ }\textbf {\bibinfo {volume} {130}} (\bibinfo {year}
  {2009}),\ 10.1063/1.3126249},\ \bibinfo {note} {184103}\BibitemShut {NoStop}%
\bibitem [{\citenamefont {Salihbegovic}(2023)}]{github}%
  \BibitemOpen
  \bibfield  {author} {\bibinfo {author} {\bibfnamefont {F.}~\bibnamefont
  {Salihbegovic}},\ }\href {https://github.com/salihbegovic/Silicon-CCSD-T-}
  {\enquote {\bibinfo {title} {Silicon interstitials ccsd(t)},}\ } (\bibinfo
  {year} {2023})\BibitemShut {NoStop}%
\bibitem [{\citenamefont {{Salihbegovic}}(2023)}]{zenodo}%
  \BibitemOpen
  \bibfield  {author} {\bibinfo {author} {\bibnamefont {{Salihbegovic}}},\
  }\href {\doibase 10.5281/ZENODO.8059575} {\enquote {\bibinfo {title}
  {salihbegovic/silicon-ccsd-t-: Workflow si-interstitials ccsd(t)},}\ }
  (\bibinfo {year} {2023})\BibitemShut {NoStop}%
\bibitem [{\citenamefont {Schäfer}\ \emph {et~al.}(2021)\citenamefont
  {Schäfer}, \citenamefont {Libisch}, \citenamefont {Kresse},\ and\
  \citenamefont {Grüneis}}]{embedding}%
  \BibitemOpen
  \bibfield  {author} {\bibinfo {author} {\bibfnamefont {T.}~\bibnamefont
  {Schäfer}}, \bibinfo {author} {\bibfnamefont {F.}~\bibnamefont {Libisch}},
  \bibinfo {author} {\bibfnamefont {G.}~\bibnamefont {Kresse}}, \ and\ \bibinfo
  {author} {\bibfnamefont {A.}~\bibnamefont {Grüneis}},\ }\href {\doibase
  10.1063/5.0036363} {\bibfield  {journal} {\bibinfo  {journal} {The Journal of
  Chemical Physics}\ }\textbf {\bibinfo {volume} {154}},\ \bibinfo {pages}
  {011101} (\bibinfo {year} {2021})}\BibitemShut {NoStop}%
\end{thebibliography}%
\clearpage

\end{document}


\title{%
  Supporting information: Formation energies of silicon
  self-interstitials using periodic coupled cluster theory
}
\author{Faruk Salihbegovi\'c}
\author{Alejandro Gallo}
\author{Andreas Grüneis}
\affiliation{%
  Institute for Theoretical Physics, Vienna University of Technology
  (TU Wien).  A-1040 Vienna, Austria, EU.
}

\maketitle

\section{\( \Gamma \)-point calculations and basis set convergence}

Table~\ref{table1} shows the \gls{ccsd} and \gls{ccsdt} energy
corrections to the \gls{hf} ground state energies, as well as the
basis set and the finite size correction.  The \gls{hf} energy was
calculated using a $\Gamma$-centered 7$\times$7$\times$7 $k$-point
mesh.  All the other corrections were calculated at the
$\Gamma$-point.

\begin{table}[h]
  \caption{%
    \gls{hf} formation energies of all calculated
    structures, as well as the \gls{ccsd}, \gls{ccsdt}, finite size
    and basis set corrections. All energies are in eV.\label{table1}
  }
    \begin{tabular}{lccccc}
      \toprule
      \gls{hf}  & $N_{\mathrm{v}}/N_{\mathrm{occ}}$ & \gls{ccsd} & \gls{ccsdt} & FS      & BS       \\
      \midrule
      Bulk      & 5                                 & -43.4765   & -2.3715     & -3.7478 & -9.7135  \\
      -151.920  & 10                                & -49.9044   & -3.4427     & -3.8412 & -4.0339  \\
                & 15                                & -51.6199   & -3.7680     & -3.8593 & -2.4171  \\
                & 20                                & -52.3780   & -3.9174     & -3.8633 & -1.6891  \\
                & 25                                & -52.8151   & -4.0039     & -3.8653 & -1.2599  \\
                & 30                                & -53.0515   & -4.0461     & -3.8654 & -1.0200  \\
      \midrule
      C3V       & 5                                 & -48.0369   & -3.1975     & -4.4917 & -10.6451 \\
      -152.9134 & 10                                & -55.0409   & -4.3890     & -4.5530 & -4.4268  \\
                & 15                                & -56.9736   & -4.7637     & -4.5677 & -2.6377  \\
                & 20                                & -57.8062   & -4.9321     & -4.5715 & -1.8250  \\
                & 25                                & -58.2881   & -5.0274     & -4.5724 & -1.3807  \\
                & 30                                & -58.5542   & -5.0770     & -4.5723 & -1.0903  \\
      \midrule
      X         & 5                                 & -48.2597   & -3.2894     & -4.5609 & -10.5387 \\
      -153.4851 & 10                                & -55.1774   & -4.4519     & -4.6033 & -4.3915  \\
                & 15                                & -57.0905   & -4.8230     & -4.6169 & -2.5965  \\
                & 20                                & -57.9043   & -4.9867     & -4.6201 & -1.8062  \\
                & 25                                & -58.3773   & -5.0805     & -4.6210 & -1.3409  \\
                & 30                                & -58.6339   & -5.1316     & -4.6207 & -1.0757  \\
      \midrule
      T         & 5                                 & -48.2917   & -3.3253     & -4.6008 & -10.7425 \\
      -151.4606 & 10                                & -55.3498   & -4.5223     & -4.6540 & -4.4593  \\
                & 15                                & -57.2692   & -4.8940     & -4.6676 & -2.6304  \\
                & 20                                & -58.0948   & -5.0627     & -4.6716 & -1.8590  \\
                & 25                                & -58.5816   & -5.1593     & -4.6727 & -1.3693  \\
                & 30                                & -58.8459   & -5.2094     & -4.6727 & -1.1039  \\
      \midrule
      H         & 5                                 & -47.9977   & -3.1610     & -4.4635 & -10.6933 \\
      -153.2528 & 10                                & -54.9550   & -4.3508     & -4.5291 & -4.4641  \\
                & 15                                & -56.8764   & -4.7222     & -4.5437 & -2.6421  \\
                & 20                                & -57.7108   & -4.8892     & -4.5467 & -1.8522  \\
                & 25                                & -58.1881   & -4.9839     & -4.5476 & -1.3534  \\
                & 30                                & -58.4548   & -5.0331     & -4.5472 & -1.1027  \\
      \midrule
      V         & 5                                 & -41.0219   & -2.6484     & -3.9846 & -9.3988  \\
      -136.8709 & 10                                & -47.3057   & -3.7151     & -4.0613 & -3.8637  \\
                & 15                                & -48.9880   & -4.0494     & -4.0775 & -2.2835  \\
                & 20                                & -49.7164   & -4.2020     & -4.0798 & -1.6001  \\
                & 25                                & -50.1419   & -4.2861     & -4.0802 & -1.1794  \\
                & 30                                & -50.3710   & -4.3320     & -4.0803 & -0.9470  \\
      \bottomrule
    \end{tabular}
\end{table}

\newpage
\section{Random \( k \)-point calculations}
Tables~\ref{bulk}-\ref{V} show the \gls{ccsd} and \gls{ccsdt} energy
corrections to the \gls{hf} ground state energies, as well as the
basis set and the finite size correction for 10 random $k$-points
using 10 virtual orbitals per occupied orbital.

\begin{table}[h]
\begin{minipage}{0.4\textwidth}
  \caption{All energies are in eV.\label{bulk}}
    \begin{tabular}{lcccc}
      \toprule
      \gls{hf} & \gls{ccsd} & \gls{ccsdt} & FS      & BS      \\
      \midrule
      Bulk     & -49.8167   & -3.7600     & -4.2053 & -4.0146 \\
      -151.920 & -49.7882   & -3.6116     & -4.0349 & -4.0027 \\
               & -49.7887   & -3.6936     & -4.1435 & -4.0076 \\
               & -49.7851   & -3.6124     & -4.0151 & -4.0076 \\
               & -49.7881   & -3.6590     & -4.1462 & -4.0077 \\
               & -49.8101   & -3.7460     & -4.1749 & -4.0190 \\
               & -49.7918   & -3.6290     & -4.0086 & -4.0090 \\
               & -49.8083   & -3.7545     & -4.2244 & -4.0303 \\
               & -49.8207   & -3.7633     & -4.1717 & -4.0173 \\
               & -49.7976   & -3.7073     & -4.1158 & -4.0102 \\
      \midrule
      Average  & -49.7995   & -3.6937     & -4.1240 & -4.0126 \\
      \bottomrule
    \end{tabular}
\end{minipage}
\begin{minipage}{0.4\textwidth}
  \caption{All energies are in eV.\label{C3V}}
    \begin{tabular}{lcccc}
      \toprule
      \gls{hf}  & \gls{ccsd} & \gls{ccsdt} & FS       & BS       \\
      \midrule
      C3V       & -55.0793   & -4.6865     & -4.8126  & -4.3949  \\
      -152.9134 & -55.0242   & -4.5504     & -4.6772  & -4.3986  \\
                & -55.1168   & -4.6773     & -4.8237  & -4.3960  \\
                & -55.0566   & -4.5760     & -4.6621  & -4.3849  \\
                & -55.0991   & -4.6385     & -4.8578  & -4.3967  \\
                & -55.1570   & -4.7165     & -4.7976  & -4.3899  \\
                & -55.0867   & -4.6046     & -4.6337  & -4.3900  \\
                & -55.1814   & -4.7321     & -4.9133  & -4.3893  \\
                & -55.1799   & -4.7372     & -4.7677  & -4.3942  \\
                & -55.1135   & -4.6757     & -4.7427  & -4.3928  \\
      \midrule
      Average   & -55.1094   & -4.65948    & -4.76885 & -4.39272 \\
      \bottomrule
    \end{tabular}
\end{minipage}
\end{table}

\begin{table}[h]
\begin{minipage}{0.4\textwidth}
  \caption{All energies are in eV.\label{X}}
    \begin{tabular}{lcccc}
      \toprule
      \gls{hf}  & \gls{ccsd} & \gls{ccsdt} & FS       & BS       \\
      \midrule
      X         & -55.0427   & -4.6651     & -4.8119  & -4.3681  \\
      -153.4851 & -55.1184   & -4.5860     & -4.7078  & -4.3438  \\
                & -55.0224   & -4.6324     & -4.7803  & -4.3883  \\
                & -55.0596   & -4.5699     & -4.6657  & -4.3641  \\
                & -55.0501   & -4.6117     & -4.8226  & -4.3873  \\
                & -55.0537   & -4.6725     & -4.8059  & -4.3848  \\
                & -55.0422   & -4.5854     & -4.6454  & -4.3788  \\
                & -55.0168   & -4.6710     & -4.9005  & -4.4082  \\
                & -55.0689   & -4.6791     & -4.7565  & -4.3684  \\
                & -55.0544   & -4.6451     & -4.7333  & -4.3727  \\
      \midrule
      Average   & -55.0530   & -4.63182    & -4.76299 & -4.37646 \\
      \bottomrule
    \end{tabular}
\end{minipage}
\begin{minipage}{0.4\textwidth}
  \caption{All energies are in eV.\label{T}}
    \begin{tabular}{lcccc}
      \toprule
      \gls{hf}  & \gls{ccsd} & \gls{ccsdt} & FS       & BS       \\
      \midrule
      T         & -55.2699   & -4.7589     & -4.8835  & -4.3863  \\
      -151.4606 & -55.2986   & -4.6747     & -4.7801  & -4.3943  \\
                & -55.3547   & -4.7756     & -4.8900  & -4.3931  \\
                & -55.2511   & -4.6547     & -4.7454  & -4.4002  \\
                & -55.1553   & -4.6366     & -4.8288  & -4.3856  \\
                & -55.2156   & -4.7206     & -4.8375  & -4.3851  \\
                & -55.0346   & -4.5578     & -4.6527  & -4.4022  \\
                & -55.0616   & -4.6681     & -4.8444  & -4.4209  \\
                & -55.1094   & -4.6811     & -4.7886  & -4.3854  \\
                & -55.1468   & -4.6625     & -4.7681  & -4.3789  \\
      \midrule
      Average   & -55.1897   & -4.67906    & -4.80191 & -4.39320 \\
      \bottomrule
    \end{tabular}
\end{minipage}
\end{table}

\begin{table}[h]
\begin{minipage}{0.4\textwidth}
  \caption{All energies are in eV.\label{H}}
    \begin{tabular}{lcccc}
      \toprule
      \gls{hf}  & \gls{ccsd} & \gls{ccsdt} & FS       & BS       \\
      \midrule
      H         & -55.0364   & -4.6706     & -4.7986  & -4.3909  \\
      -153.2528 & -54.9589   & -4.5200     & -4.6609  & -4.3890  \\
                & -55.0356   & -4.6364     & -4.7915  & -4.3832  \\
                & -54.9648   & -4.5345     & -4.6405  & -4.3797  \\
                & -55.0054   & -4.5977     & -4.8300  & -4.3916  \\
                & -55.0608   & -4.6697     & -4.7742  & -4.3860  \\
                & -54.9979   & -4.5684     & -4.6153  & -4.3882  \\
                & -55.0974   & -4.6879     & -4.8865  & -4.3808  \\
                & -55.0909   & -4.6923     & -4.7437  & -4.3858  \\
                & -55.0385   & -4.6379     & -4.7173  & -4.3803  \\
      \midrule
      Average   & -55.0287   & -4.62153    & -4.74585 & -4.38555 \\
      \bottomrule
    \end{tabular}
\end{minipage}
\begin{minipage}{0.4\textwidth}
  \caption{All energies are in eV.\label{V}}
    \begin{tabular}{lcccc}
      \toprule
      \gls{hf}  & \gls{ccsd} & \gls{ccsdt} & FS       & BS       \\
      \midrule
      V         & -46.8932   & -3.7864     & -4.1677  & -3.8147  \\
      -136.8709 & -46.9264   & -3.7007     & -4.0489  & -3.8173  \\
                & -46.9694   & -3.7560     & -4.1327  & -3.7825  \\
                & -46.8965   & -3.6820     & -4.0553  & -3.8076  \\
                & -46.8375   & -3.6923     & -4.0976  & -3.8153  \\
                & -46.8939   & -3.7783     & -4.1499  & -3.8024  \\
                & -46.8376   & -3.6612     & -4.0657  & -3.8129  \\
                & -46.7661   & -3.7462     & -4.1541  & -3.8628  \\
                & -46.8131   & -3.7611     & -4.1651  & -3.8299  \\
                & -46.8483   & -3.7296     & -4.1263  & -3.8017  \\
      \midrule
      Average   & -46.8682   & -3.72940    & -4.11633 & -3.81470 \\
      \bottomrule
    \end{tabular}
\end{minipage}
\end{table}
